\documentclass[twocolumn]{aastex631}

\usepackage[version=3]{mhchem}
\usepackage{multirow}

\usepackage{xcolor, soul}
\sethlcolor{green}

\begin{document}

\title{Infrared Spectroscopy and Photochemistry of Aromatic Nitriles in Para-Hydrogen Matrices}

\author{Sam McGrath}
\affiliation{Department of Chemistry, University of British Columbia, 2036 Main Mall, Vancouver BC V6T 1Z1, Canada}
\author[0000-0001-6035-3869]{Vincent J. Esposito}
\affiliation{Schmid College of Science and Technology, Chapman University, Orange, California 92866, United States}
\author{Linshan Zeng}
\affiliation{Department of Chemistry, University of British Columbia, 2036 Main Mall, Vancouver BC V6T 1Z1, Canada}
\author[0000-0001-8134-5681]{Thomas H. Speak}
\affiliation{Department of Chemistry, University of British Columbia, 2036 Main Mall, Vancouver BC V6T 1Z1, Canada}
\author{Brendan Moore}
\affiliation{Department of Chemistry, University of British Columbia, 2036 Main Mall, Vancouver BC V6T 1Z1, Canada}
\author{Pavle Djuricanin}
\affiliation{Department of Chemistry, University of British Columbia, 2036 Main Mall, Vancouver BC V6T 1Z1, Canada}
\author[0000-0003-1951-8515]{Jun Miyazaki}
\affiliation{Department of Natural Sciences, Tokyo Denki University, 5 Senju-Asahi-cho, Adachi-ku, Tokyo 120-8551, Japan}
\author[0000-0001-8976-1938]{Takamasa Momose}
\affiliation{Department of Chemistry, University of British Columbia, 2036 Main Mall, Vancouver BC V6T 1Z1, Canada}
\author[0000-0002-0850-7426]{Ilsa R. Cooke}
\affiliation{Department of Chemistry, University of British Columbia, 2036 Main Mall, Vancouver BC V6T 1Z1, Canada}

\correspondingauthor{Ilsa R. Cooke}
\email{icooke@chem.ubc.ca}

\begin{abstract}

Motivated by recent detections of several aromatic nitriles in Taurus Molecular Cloud-1, we report laboratory and theoretical investigations of the vibrational spectroscopy and photochemistry of singly and doubly cyano-substituted benzene in solid para-hydrogen matrices. We compare the photochemistry of cyanobenzene (benzonitrile) and three dicyanobenzene isomers initiated by excitations at 193 nm. In addition, we report the photochemistry of deuterated cyanobenzene (d$_5$-cyanobenzene), enabling us to determine the major products produced during the cyanobenzene photodissociation. The major products observed in the photolysis of all the nitriles are HCN and HNC, which are likely produced by hydrogen abstraction from para-H$_2$ by the CN radical. This indicates that the major photodissociation channel involves cleavage of the bond between the ring and the nitrile group, forming the phenyl (or cyanophenyl) radical + CN. We observe secondary photoproducts similar to those found during benzene photolysis. Our findings may aid the interpretation of recent JWST mid-infrared observations of aromatics in photodissociation regions. 

\end{abstract}

\keywords{Interstellar medium (847) --- Dense interstellar clouds (371) --- Interstellar molecules (849) --- Astrochemistry (75) --- Polycyclic aromatic hydrocarbons (1280) --- Laboratory astrophysics (2004) }

\section{Introduction} \label{sec:intro}

Aromatic molecules are ubiquitous structural motifs, not only in the chemical makeup of life, but also in the broader chemical evolution of the universe. Analysis of the so-called unidentified infrared bands (UIRs) suggests that as much as 25\% of all interstellar carbon is thought to be locked up in the form of polycyclic aromatic hydrocarbons (PAHs; \citealt{Chiar2013}). These molecules are traditionally thought to form primarily, if not exclusively, in the circumstellar envelopes of evolved stars \citep{Tielens:2008}, from which they are ejected into the diffuse medium. It has been suggested that only large PAHs ($>$30--40 carbon atoms) can survive their journey through the harsh conditions of the diffuse medium to the dense cloud stage \citep{chabot_coulomb_220,BETTENS1995321}. The presence of several small aromatic molecules in the dark cloud core TMC-1 (Taurus Molecular Cloud-1 ), far removed from the envelope of an evolved star, was therefore unexpected. 

The cyanopolyyne peak of TMC-1 (TMC-1 CP) is one of the most well-studied dense cores, in which an abundance of complex organic molecules (COMs) have been observed, including carbon-chains, cations and anions, oxygen-bearing species and nitriles \citep{Gratier:2016fj}. More recently, the discovery of benzonitrile (c-C$_6$H$_5$CN) by \citet{McGuire:2018it} added the first aromatic ring to this inventory. Following this detection, several additional aromatic and unsaturated cyclic nitriles have been detected in the GOTHAM (GBT Observations of TMC-1: Hunting for Aromatic Molecules) line survey \citep{McCarthy:2021aa,McGuire2021,Burkhardt2021,Lee:2021ud,Sita2022,wenzel2024,wenzel2025,wenzel2025c} and the QUIJOTE (Q-band Ultrasensitive Inspection Journey to the Obscure TMC-1 Environment) survey \citep{Cernicharo2021,Cernicharo2021a,Cernicharo2021b, Cernicharo2022,cernicharo2024,cabezas2025,cernicharo2026,fuentetaja2026}. 

Despite their apparent ubiquity, the formation and evolution of PAHs remain poorly constrained, highlighting the need for quantitative measurements of formation and destruction routes relevant to dense molecular clouds. It has been suggested that a bottleneck to the formation of these large aromatics at low temperatures is the cyclization to produce the first aromatic ring, usually benzene or the phenyl radical \citep{Cherchneff1992,Tielens1997,Kaiser2015}. 
However, the detection of benzene in interstellar space is challenging as it does not have a permanent dipole moment, meaning it cannot be observed using radio telescopes. Benzene has been detected in a handful of sources through observations of the $\nu_4$ bending mode; however, these observations require bright IR background stars, making it challenging to observe benzene in dense cores \citep{Cernicharo2001,Kraemer2006,Malek2011}. One way that has been proposed to circumvent this is to use cyanobenzene (benzonitrile) as a chemical proxy for benzene in order to determine its abundance in dense molecular clouds. Cyanobenzene has been shown to form through a barrierless exothermic neutral-neutral reaction between benzene and CN \citep{Woon:2006ce,Trevitt2010, Cooke:2020we}, and therefore is a viable reaction under cold cloud conditions. While a range of aromatic nitriles have now been detected in dense clouds, their interstellar life cycle and transport during star formation is not well understood

% Within cold dark clouds, such as TMC-1, dust grains act as a reaction site for the formation of COMs. These dust grains can stabilize intermediates that would not be stabilized in the gas phase by absorbing and transferring energy produced during reactions. It has been argued that COMs can form in these ices, followed by their desorption from the ice as the temperature of the cloud begins to rise during the protostellar phase or by non-thermal mechanisms. 

In order to understand the lifetime of these molecules in interstellar space and their survival to the protoplanetary disk phase, it is essential to understand their photochemistry. Molecules within the interiors of dense clouds are shielded from the external interstellar radiation field. However, cosmic rays can penetrate deep into these clouds, producing secondary electrons that excite H$_2$, which subsequently relaxes by releasing VUV photons with strong emissions at the Lyman-$\alpha$ wavelength, 121.567 nm \citep{Gredel1989}. This VUV radiation interacts with molecules in both the gas phase and in ice, resulting in their excitation, dissociation, isomerization, or photodesorption from ice surfaces \citep{Oeberg2016}. 

Matrix isolation is a technique to trap molecules within a nonreactive matrix crystal. Solid parahydrogen (p-H$_2$) matrix isolation provides a soft environment to monitor the dissociation behaviour of astrochemical molecules in an environment analogous to the gas phase at low temperatures. Unlike traditional noble gas matrices such as argon, neon, and nitrogen, solid para-hydrogen is free of the ``cage effect'', which prevents the matrix environment from disturbing the photodissociation process \citep{Fajardo2013, Huang2010}. 

\begin{figure}[h!]
\centering
\includegraphics[width=0.45\textwidth]{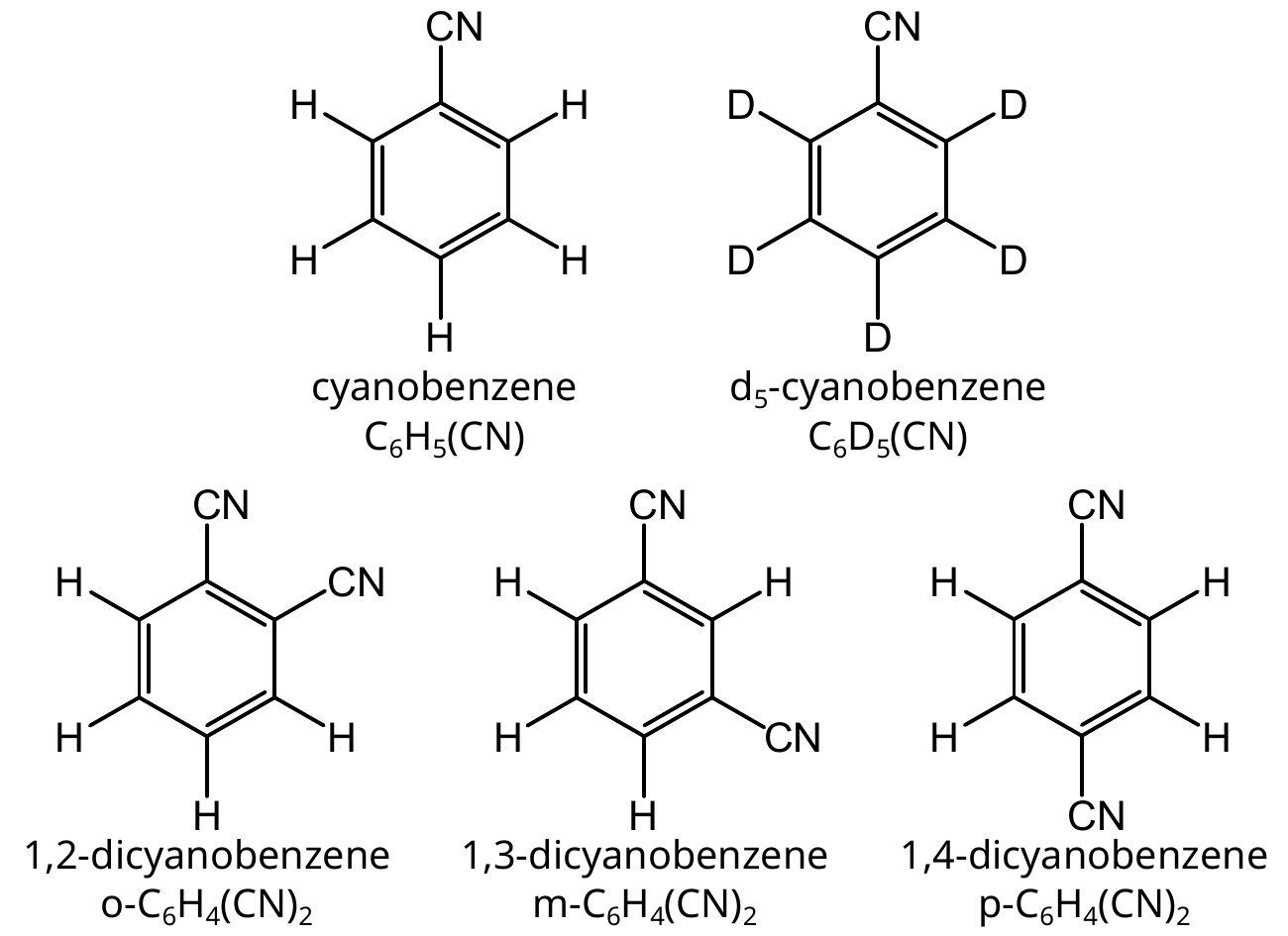}
\caption{Chemical structure of the aromatic nitriles cyanobenzene, d$_5$-cyanobenzene, 1,2-dicyanobenzene, 1,3-dicyanobenzene and 1,4-dicyanobenzene}
\label{fig:molecules}
\end{figure}

We have measured and assigned the mid-infrared spectra of the aromatic nitriles cyanobenzene and 1,2- (ortho), 1,3-(meta), and 1,4-(para) dicyanobenzene (Fig. \ref{fig:molecules}) embedded in para-hydrogen matrices. We have characterized the photoproducts following the 193 nm irradiation of these molecules in conditions similar to those of interstellar space. 

\section{Methods} \label{sec:exp}

\subsection{Experimental}
The experiments were conducted using the setup described previously in \citet{Toh2015}. Briefly, the apparatus is composed of a para-hydrogen converter, a high-vacuum chamber, a high-resolution Fourier transform infrared (FTIR) spectrometer, and a photolysis source (either a 193 nm ArF excimer laser or a 213 nm Nd:YAG laser).
Parahydrogen was prepared by a similar method as mentioned in \citet{Miyamoto2012} and \citet{Tom2009}.  Briefly, normal hydrogen gas of 99.99\% purity (Praxair Canada Inc.) was converted into parahydrogen gas, with a purity of 99.95\%, by flowing through a magnetic catalyst (FeOH)O kept at a temperature of about 14 K. The parahydrogen was stored in stainless steel cylindrical containers before entering the vacuum chamber \citep{Tom2009}.

The inner matrix chamber has a cesium iodide (CsI) plate fixed in the center of a nickel-plated copper plate holder, which is cooled to 3.7 K by a closed-cycle helium refrigerator (Sumitomo Heavy Industries, Ltd., SRDK-205). Liquid samples (benzene, cyanobenzene, d$_5$-cyanobenzene) were prepared by pre-filling into a sample cylinder, whereas solid samples (1,2-dicyanobenzene, 1,3-dicyanobenzene and 1,4-dicyanobenzene) were deposited via a Knudsen cell located in the matrix chamber. A thin layer of pure parahydrogen was co-deposited onto the crystal to prevent sample evaporation.

Cyanobenzene (Sigma-Aldrich, 99\%) and d$_5$-cyanobenzene (Sigma-Aldrich, 99\%) were purified by freeze-pump-thaw cycles at 77 K and room temperature.  1,2-dicyanobenzene (Sigma-Aldrich, 98\%), 1,3-dicyanobenzene (Sigma-Aldrich, 98\%) and 1,4-dicyanobenzene (Sigma-Aldrich, 98\%) were directly deposited into the Knudsen cell without purification. All sample crystals after deposition were annealed at 4.7 K for 1 hour before exposure to the photolysis laser, to prevent sample evaporation on the surface. An ArF excimer laser (193 nm, CEX-100, 5 ns) and the fifth harmonic of an Nd:YAG laser (213 nm, Litron, Nano-SG 150-10, 4 ns) were used as radiation sources.

Infrared absorption spectra were taken with a high-resolution FTIR (Bruker, IFS 125HR) equipped with a KBr beam splitter, a glowbar MIR light source, and a liquid N$_2$ cooled MCT detector. All spectra were recorded at a resolution of 0.1-0.2 cm$^{-1}$ and averaged over 1000 scans. For kinetic measurements, FTIR scans were taken after the deposition and annealing of the para-H$_2$ matrix and successive scans were taken in 150~s intervals following irradiation. The laser power was measured and used to determine the laser fluence each sample was exposed to.

\subsection{Computational}
Two different methods are utilized to compute the anharmonic vibrational spectra of the molecules in this study. For cyanobenzene, the computations begin with a geometry optimization followed by calculation of the normal mode vectors and harmonic vibrational frequencies using the B3LYP method and the N07D basis set \citep{Barone2008}. Following this, a quartic force field (QFF) is computed at the same level of theory. A quartic force field is a fourth-order Taylor expansion of the potential energy surface around the equilibrium geometry. Single-point energies are computed at points around the equilibrium geometry at defined displacements to determine the quadratic, cubic, and quartic force constants. A linear transformation is used to transform the QFF from normal to Cartesian coordinates.

Second-order vibrational perturbation theory (VPT2) is then used to compute the anharmonic vibrational frequencies via a modified version of the \textsc{SPECTRO} software \citep{SPECTRO}. \textsc{SPECTRO} implements resonance polyad matrices in the VPT2 treatment. These matrices allow for a more advanced treatment of vibrational resonances by simultaneously capturing the resonance interactions. The polyad matrices are built based on mode symmetry and a frequency separation threshold, which is set to 200~cm$^{-1}$ here. The resonance polyad approach also allows for intensity redistribution based on the coupling strength between the vibrational modes, a process that is difficult to describe with standard VPT2 models. 

For the three isomers of dicyanobenzene, a similar QFFare approach is used with some modifications. In this instance, geometry optimization, normal mode determination, and calculation of the harmonic frequencies is performed using the rev-DSDPBEP86 (rDSD) double-hybrid density functional in conjunction with the jun-cc-pVTZ (TZ) basis set. The MP2 electron correlation included in this method has been shown to better treat the electronic structure of cyano-substituted PAHs \citep{Esposito2025,Fortenberry2026}. 

Following this, the cubic and quartic force constants are computed with the B3LYP/N07D level of theory. Combination of the rDSD/TZ quadratic force constants with the cubic and quartic force constants from B3LYP/N07D creates a hybrid QFF. VPT2 is then used to compute the anharmonic vibrational spectra with \textsc{SPECTRO} and the same threshold conditions. This hybrid QFF approach is also used to compute the CN stretch of cyanobenzene.

\section{Results}
\tabletypesize{\footnotesize}
\begin{deluxetable}{ccccc}[hbt!]
\tablecaption{Measured IR transitions in cm$^{-1}$, anharmonic computed vibrational transitions in cm$^{-1}$, gas-phase values from the literature, the corresponding vibrational mode assignments and relative intensities for cyanobenzene in solid para-H$_{2}$. Anharmonic data are computed at the B3LYP/N07D level of theory. }
\label{table:cyano_dep}
\tablehead{ 
\multicolumn{5}{c}{\centering Cyanobenzene} \\ 
\colhead{ID}&
\colhead{p-H$_2$ (I)}&
\colhead{Calc. (I)}& 
\colhead{Gas*}&
\colhead{Mode}}
\startdata
a & 545.3 (0.18) & 556.7 (0.32) & 547 & $\nu_{27}$ \\
b & 687.8 (0.59) & 688.9 (0.66) & 687 & $\nu_{29}$ \\ 
c & 757.6 (1.00) & 760.6 (1.00) & 758 & $\nu_{24}$ \\
  & 757.9 (1.00)  & &  & \\
d & 923.0 (0.08) & 923.7 (0.07) & 926 & $\nu_{21}$ \\
e & 1028.8 (0.09) & 1030.1 (0.09) & 1027 & $\nu_{17}$ \\
f & 1070.8 (0.06) & 1075.3 (0.06) & 1071 & $\nu_{16}$ \\
g & 1095.7 (0.03) & 1101.8 (0.01) &  & $\nu_{25}$+$\nu_{30}$/$\nu_{16}$ \\
h & 1161.6 (0.02) & 1173.2 (0.01) & 1163 & $\nu_{15}$ \\
i & 1179.5 (0.04) & 1184.6 (0.02) & 1178 & $\nu_{14}$ \\
j & 1193.3 (0.01) & 1198.4 (0.01) & 1193 & $\nu_{13}$ \\
k & 1286.4 (0.05) & 1300.9 (0.01) & 1288 & $\nu_{12}$ \\
l & 1333.1 (0.02) & 1336.8 (0.01) & 1335 & $\nu_{11}$ \\
m & 1450.5 (0.17) & 1449.2 (0.12) & 1448 & $\nu_{10}$ \\
n & 1495.0 (0.29) & 1494.1 (0.12) & 1491 & $\nu_{9}$ \\
o & 1530.9 (0.02) & 1533.0 (0.01) & & $\nu_{22}$+$\nu_{25}$ \\
p & 1585.8 (0.04) & 1578.5 (0.01) & 1583 & $\nu_{8}$ \\
q & 1600.2 (0.04) & 1603.9 (0.02) & 1599 & $\nu_{22}$+$\nu_{24}$ \\
  & 1601.0 (0.04) &  & & \\
r & 1680.6 (0.04) & 1688.2 (0.02) & & $\nu_{21}$+$\nu_{24}$ \\
s & 1764.5 (0.04) & 1773.3 (0.01) &  & $\nu_{21}$+$\nu_{22}$ \\
t & 1811.8 (0.05) & 1825.4 (0.03) & & $\nu_{20}$+$\nu_{22}$ \\
u & 1891.9 (0.05) & 1891.3 (0.04) & & $\nu_{20}$+$\nu_{21}$ \\
v & 1911.7 (0.03) & 1911.6 (0.04) & & $\nu_{19}$+$\nu_{21}$ \\
w & 1959.4 (0.05) & 1957.8 (0.04) & & $\nu_{19}$+$\nu_{20}$ \\
  & 1959.8 (0.05) & & &  \\
x & 1979.4 (0.03) &  & & \\
  & 1979.8 (0.03) & & &  \\
y & 2238.2 (0.38) & 2225.8 (0.84) & 2229 & $\nu_{6}$\\ 
  & 2241.9 (0.38) &  & &  \\ 
z & 3010.0 (0.01) & 3005.2 (0.01) & & $\nu_{8}$+$\nu_{10}$\\
A & 3037.8 (0.05) & 3031.3 (0.04) & 3027 & $\nu_{7}$+$\nu_{10}$ \\
B & 3049.4 (0.05) & 3047.7 (0.09) & 3043 & $\nu_{3}$ \\
C & 3064.0 (0.02) & 3061.9 (0.05) &  & $\nu_{8}$+$\nu_{9}$ \\
D & 3069.0 (0.11) & 3072.9 (0.07) & 3066 & $\nu_{4}$ \\
E & 3073.0 (0.05) & 3076.3 (0.07) &  & $\nu_{3}$/$\nu_{5}$ \\ 
F & 3079.6 (0.10) & 3084.8 (0.05) &  & $\nu_{1}$ \\
G & 3098.2 (0.10) & 3106.4 (0.09) & 3093 & $\nu_{2}$ \\
H & 3113.8 (0.03) & 3120.1 (0.07) & 3106 & $\nu_{7}$+$\nu_{9}$ \\
I & 3168.9 (0.01) & 3167.6 (0.01) &  & 2$\nu_{8}$ 
\enddata
\tablenotetext{a}{The anharmonic transitions presented in Figure~\ref{fig:cn_comparison} are computed at the rDSD/junTZ+B3LYP/N07D level of theory. See text for further discussion.} 
\end{deluxetable}

\subsection{Cyanobenzene spectroscopy}

Infrared spectra of cyanobenzene and d$_5$-cyanobenzene in solid p-H$_2$ were recorded prior to irradiation; example deposition spectra can be found in Appendix \ref{sec:dep-spectra}. Table \ref{table:cyano_dep} summarizes the measured frequency and relative intensity of each observed transition of cyanobenzene in the p-H$_2$ matrix. These values were then compared with theoretical calculations and prior gas-phase studies \citep{Rajasekhar2022} for assignment of the vibrational modes. The intensities are given relative to the most intense band observed, which is at 757.6 cm$^{-1}$ for cyanobenzene. A table summarizing the observed vibrations for d$_5$-cyanobenzene in p-H$_2$ and a comparison with calculated values can be found in the Appendix Table~\ref{table:d_cyano_dep}. Below, we describe the major features observed in the spectra and their assignment. 

\textbf{500-2000 cm$^{-1}$:} The lowest observable peak at 545.3 cm$^{-1}$, assigned to $\nu_{27}$, shows a frequency shift of +1.7 cm$^{-1}$ from the gas phase and a difference of 11.4 cm$^{-1}$ from theoretical calculations. The peak at 687.8 cm$^{-1}$ is assigned to $\nu_{29}$, shifted +0.8 cm$^{-1}$ from the gas phase, and a difference of 1.1 cm$^{-1}$ from theoretical calculations.  
The most intense absorption appears as a doublet in p-H$_2$, at 757.6  cm$^{-1}$ and 757.9  cm$^{-1}$. It is assigned to $\nu_{24}$, with a slight shift of -0.4/0.1 cm$^{-1}$ from the gas phase and a difference of 3.3/3.0 cm$^{-1}$ from theoretical calculations.
The peaks at 1450.5 cm$^{-1}$ and 1495.0 cm$^{-1}$ are assigned to modes $\nu_{10}$ and $\nu_{9}$, respectively. Both modes involve stretching of the CC bonds within the aromatic ring. 

\textbf{CN stretch, $\nu$(CN):} For cyanobenzene, two peaks are observed at 2238.2 cm$^{-1}$ and 2241.1 cm$^{-1}$ as can be seen in Table \ref{table:cyano_dep} and Figure \ref{fig:cn_comparison}. These are assigned to the CN stretching mode $\nu_6$. The experimental peaks are shifted from the gas-phase value of 2229 cm$^{-1}$  \citep{Rajasekhar2022}, due to matrix effects. Two levels of theory are used to compute the CN stretch for cyanobenzene as described in the computational methods. The rDSD/TZ+B3LYP/N07D hybrid method computes a CN stretch frequency of 2225.8~cm$^{-1}$, while the B3LYP/N07D calculation produced a value of 2297 cm$^{-1}$.

\textbf{CH stretching, $\nu$(CH):} The CH stretches have been tentatively assigned in Table \ref{table:cyano_dep} due to splitting of the expected vibrational modes. Resulting from interactions with p-H$_2$ molecules in the crystal lattice and possible clustering of cyanobenzene molecules. 

\subsection{Dicyaonobenzene isomers spectroscopy}

Infrared spectra of 1,2-dicyanobenzene, 1,3-dicyanobenzene and 1,4-dicyanobenzene in solid p-H$_2$ were recorded prior to irradiation. Sample deposition spectra for each are provided in Appendix \ref{sec:dep-spectra}. 

Table \ref{table:dicyano_dep} summarises the mode labeling, measured band centers, and relative intensities of each isomer in a p-H$_2$ matrix. These values are compared with theoretical calculations for further assignment. The intensities were calculated relative to the most intense band observed, which were 766.3 cm$^{-1}$ (1,2-dicyanobenzene), 802.6 cm$^{-1}$ (1,3-dicyanobenzene), and 843.3 cm$^{-1}$ (1,4-dicyanobenzene).

\textbf{500-2000 cm$^{-1}$:} The lowest observable peak for 1,2 and 1,3-dicyanobenzene are at 522.8 cm$^{-1}$ and 497.4 cm$^{-1}$, respectively. These have been assigned to $\nu_{28}$ and show close agreement with theoretical calculations, differing only by 0.3-1.6 cm$^{-1}$. For 1,4-dicyanobenzene this peak is at 559.3 cm$^{-1}$, and is assigned to $\nu_{26}$ with a difference of 2.2 cm$^{-1}$ from theoretical calculations. The most intense vibrations in each case involve out-of-plane ring deformation and bending motions of the atoms attached to the ring. These are at 766.3 cm$^{-1}$ (1,2), 802.6 cm$^{-1}$ (1,3), and 843.3 cm$^{-1}$ (1,4), a difference from theoretical calculations in the range of 0.8--4.7 cm$^{-1}$. Each isomer also has another strongly absorbing vibrational mode, found at 1489.0 cm$^{-1}$ (1,2), 1482.6 cm$^{-1}$ (1,3), and 1505.3 cm$^{-1}$ (1,4). All of which are assigned to $\nu_{9}$, relating to CC ring stretching and CH in-plane bending motions.

\begin{figure*}[t!]
    \centering
    \includegraphics[scale =0.5]{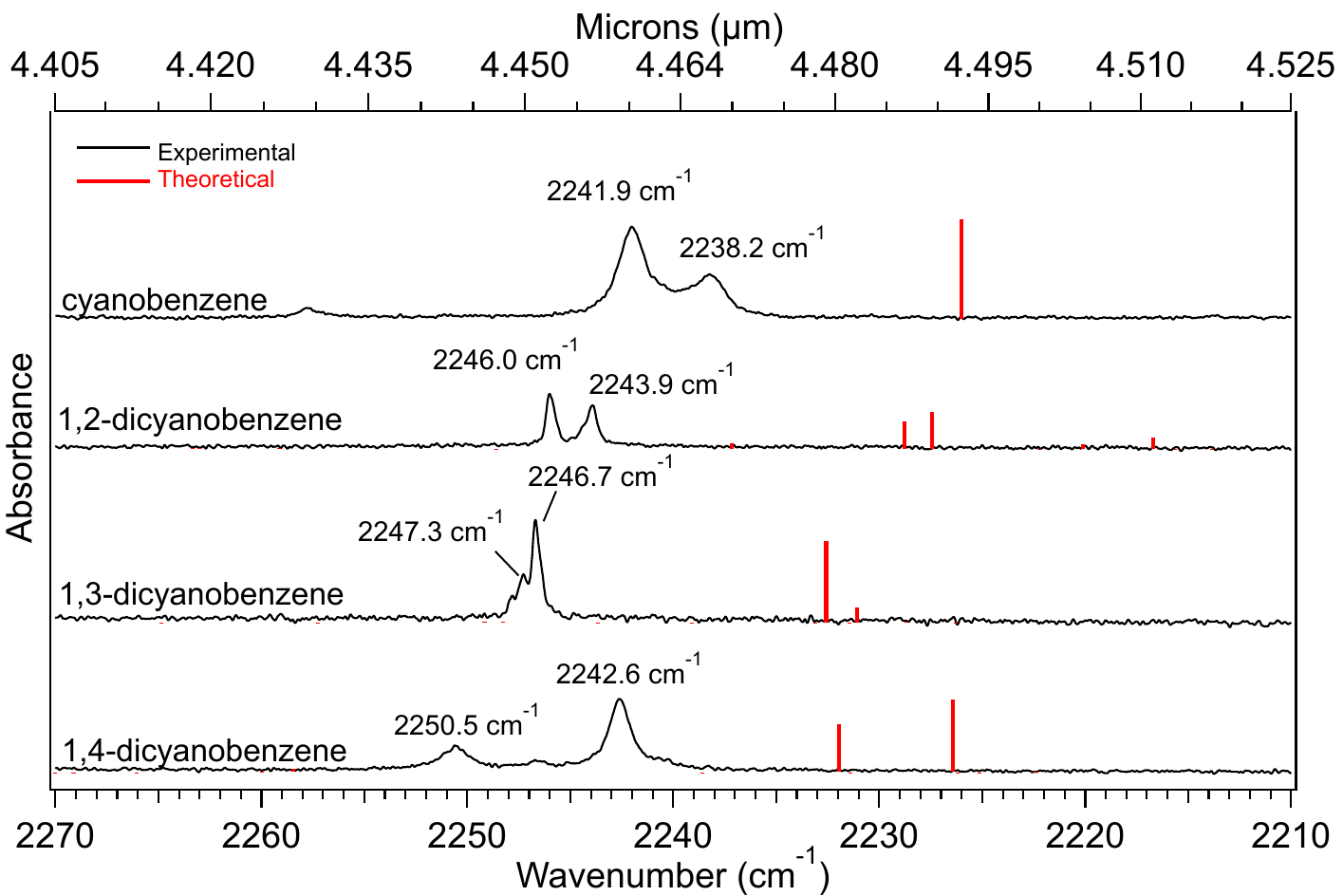}
	\caption{Comparison of the CN stretching region (shown from 2210--2270 cm$^{-1}$, or 4.405--4.525 $\mu$m) for the nitrile derivatives cyanobenzene, 1,2-dicyanobenzene, 1,3-dicyanobenzene, and 1,4-dicyanobezene, deposited in p-H$_2$ matrices.}
	\label{fig:cn_comparison}
\end{figure*}

\textbf{CN stretching, $\nu$(C-N):} There is a noticeable difference in the structure and position of the CN stretching region for each isomer, visible in Fig. \ref{fig:cn_comparison}. In the case of 1,2-dicynobenzene these peaks are observed at 2243.9 cm$^{-1}$, assigned to $\nu_{6}$, and 2246.0 cm$^{-1}$, assigned to $\nu_{5}$. For 1,3-dicyanobenzene they are found at 2246.7 cm$^{-1}$ and 2247.3 cm$^{-1}$ which are assigned to $\nu_{5}$ and $\nu_{6}$, respectively. However, the assignment is more difficult here due to overlapping peaks. Finally, these stretches are observed at 2242.6 cm$^{-1}$ and 2250.6 cm$^{-1}$ for 1,4-dicyanobenzene. Assignment is based on theoretical calculations, whose values are given in Table \ref{table:dicyano_dep}.

\textbf{CH stretching, $\nu$(C-H):}
The C-H stretches are considerably less intense for all of the dicyanobenzene isomers in comparison to cyanobenzene. For this reason, they have been tentatively assigned in Table \ref{table:dicyano_dep}. 

\begin{deluxetable*}{cccc|cccc|cccc}[b!]
\tablecaption{Measured IR transitions in cm$^{-1}$, anharmonic computed vibrational transitions in cm$^{-1}$, the corresponding vibrational mode assignments and relative intensities for 1,2-dicyanobenzene, 1,3-dicyanobenzene and 1,4-dicyanobenzene in solid para-H$_{2}$. Anharmonic data are computed at the B3LYP/N07D level of theory.}
\label{table:dicyano_dep}
\tablehead{\multicolumn{4}{c}{\textbf{1,2-dicyanobenzene}} & \multicolumn{4}{c}{\textbf{1,3-dicyanobenzene}} & \multicolumn{4}{c}{\textbf{1,4-dicyanobenzene}} \\
\cline{1-4} \cline{5-8} \cline{9-12}
\colhead{ID}&
\colhead{p-H$_2$ (I)}&
\colhead{Calc. (I)}&
\colhead{Mode}&
\colhead{ID}&
\colhead{p-$H_2$ (I)}&
\colhead{Calc. (I)}&
\colhead{Mode}&
\colhead{ID}&
\colhead{p-H$_2$ (I)}&
\colhead{Calc. (I)}&
\colhead{Mode}}
\startdata
a & 522.8 (0.29) & 523.1 (0.27) & $\nu_{28}$ & a & 497.4 (0.39) & 495.7 (0.33) & $\nu_{28}$ & a & 559.3 (0.68) & 561.5 (0.54) & $\nu_{26}$ \\
b & 766.3 (1.00) & 771.0 (1.00) & $\nu_{22}$ & b & 682.1 (0.45) & 666.7 (0.49) & $\nu_{24}$ & b & 636.6 (0.04) & 637.5 (0.01) & $\nu_{25}$ \\
c & 806.6 (0.03) & 808.5 (0.02) & $\nu_{21}$ & c & 802.6 (1.00) & 806.0 (1.00) & $\nu_{22}$ & c & 843.3 (1.00) & 853.2 (1.00) & $\nu_{20}$ \\
d & 961.0 (0.04) & 966.5 (0.03) & $\nu_{19}$ & d & 893.9 (0.12) & 898.3 (0.11) & $\nu_{21}$ & d & 1023.4 (0.14) & 1023.2 (0.07) & $\nu_{17}$ \\
e & 1037.8 (0.02) & 1043.6 (0.01) & $\nu_{17}$ & e & 907.0 (0.40) & 912.2 (0.29) & $\nu_{20}$ & e & 1107.1 (0.04) & 1120.0 (0.05) & $\nu_{16}$ \\
f & 1097.1 (0.04) & 1097.3 (0.03) & $\nu_{23}$+$\nu_{31}$/$\nu_{16}$ & f & 1098.3 (0.11) & 1101.4 (0.09) & $\nu_{16}$ & f & 1202.1 (0.02) & 1200.9 (0.01) & $\nu_{21}$+$\nu_{33}$ \\
g & 1202.6 (0.01) & 1209.6 (0.01) & $\nu_{13}$ & g & 1143.9 (0.05) & 1150.6 (0.02) & $\nu_{15}$ & g & 1261.3 (0.01) & 1263.4 (0.01) & $\nu_{23}$+$\nu_{26}$ \\
h & 1213.9 (0.01) & 1231.3 (0.01) & $\nu_{13}$ & h & 1175.5 (0.06) & 1183.4 (0.01) & $\nu_{14}$ & h & 1269.6 (0.01) & 1293.1 (0.02) & $\nu_{24}$+$\nu_{25}$ \\
i & 1225.0 (0.01) & 1241.6 (0.01) & $\nu_{20}$+$\nu_{32}$ & i & 1241.5 (0.01) & 1248.1 (0.00) & $\nu_{13}$ & i & 1275.2 (0.26) & 1307.6 (0.10) & $\nu_{14}$+$\nu_{36}$ \\
j & 1284.8 (0.04) & 1320.0 (0.04) & $\nu_{11}$& j & 1323.7 (0.04) & 1335.3 (0.03) & $\nu_{11}$ & j & 1293.3 (0.02) & 1321.3 (0.04) & $\nu_{14}$+$\nu_{35}$ \\
k & 1294.3 (0.02) & 1329.9 (0.01) & $\nu_{13}$+$\nu_{35}$ & k & 1422.8 (0.19) & 1427.4 (0.13) & $\nu_{10}$ & k & 1395.8 (0.07) & 1403.0 (0.13) & $\nu_{21}$+$\nu_{26}$/$\nu_{10}$ \\
l & 1299.5 (0.02) & 1347.1 (0.01) & $\nu_{19}$+$\nu_{31}$ & l & 1482.6 (0.52) & 1487.9 (0.30) & $\nu_{9}$ & l & 1410.9 (0.16) & 1421.8 (0.13) & $\nu_{10}$/$\nu_{21}$+$\nu_{26}$ \\
m & 1446.1 (0.07) & 1449.8 (0.07) & $\nu_{10}$ & m & 1582.9 (0.13) & 1590.1 (0.09) & $\nu_{8}$ & m & 1498.8 (0.01) & 1489.2 (0.00) & $\nu_{11}$+$\nu_{34}$ \\
n & 1453.7 (0.06) & 1458.4 (0.05) & $\nu_{17}$+$\nu_{30}$ & n & 1603.4 (0.04) & 1618.3 (0.02) & $\nu_{7}$ & n & 1499.6 (0.02) & 1492.5 (0.00) & $\nu_{16}$+$\nu_{31}$ \\
o & 1483.1 (0.02) & 1487.3 (0.06) & $\nu_{19}$+$\nu_{28}$ & o & 1729.9 (0.13) & 1739.7 (0.06) & $\nu_{19}$+$\nu_{25}$ & o & 1505.3 (0.45) & 1507.4 (0.22) & $\nu_{9}$ \\
p & 1489.0 (0.24) & 1492.9 (0.17) & $\nu_{9}$ & p & 1809.4 (0.12) & 1817.0 (0.04) & 2$\nu_{20}$ & p & 1521.5 (0.07) & 1525.4 (0.08) & $\nu_{19}$+$\nu_{26}$ \\
q & 1576.4 (0.02) & 1583.1 (0.01) & $\nu_{8}$ & q & 1854.7 (0.06) & &  & q & 1681.3 (0.08) & 1704.4 (0.04) & $\nu_{20}$+$\nu_{21}$ \\
r & 1600.9 (0.03) & 1609.3 (0.03) & $\nu_{7}$ & r & 1892.7 (0.01) & 1905.2 (0.01) & $\nu_{20}$+$\nu_{18}$ & r & 1800.7 (0.04) & 1812.9 (0.01) & $\nu_{19}$+$\nu_{20}$ \\
s & 1649.1 (0.05) & 1656.2 (0.02) & $\nu_{20}$+$\nu_{22}$ & s & 1912.1 (0.09) & 1923.1 (0.05) & $\nu_{18}$+$\nu_{19}$ & s & 1809.4 (0.03) & 1820.1 (0.03) & $\nu_{18}$+$\nu_{21}$ \\
t & 1725.5 (0.02) & 1739.7 (0.01) & $\nu_{19}$+$\nu_{22}$ & t & 1969.0 (0.07) & & & t & 1929.2 (0.10) & 1937.4 (0.07) & $\nu_{18}$+$\nu_{19}$ \\
u & 1766.8 (0.01) & 1762.6 (0.00) & $\nu_{22}$+$\nu_{18}$ & u & 2246.7 (0.26) & 2231.1 (0.03) & $\nu_{5}$ & u & 2242.6 (0.22) & 2226.4 (0.12) & $\nu_{5}$/$\nu_{14}$+$\nu_{17}$ \\
v & 1840.8 (0.03) & 1848.7 (0.03) & $\nu_{19}$+$\nu_{20}$ & v & 2247.3 (0.06) & 2232.6 (0.18) & $\nu_{6}$ & v & 2250.5 (0.08) & 2231.9 (0.08) & $\nu_{14}$+$\nu_{17}$/$\nu_{5}$ \\
w & 1874.8 (0.02) & 1880.0 (0.01) & $\nu_{18}$+$\nu_{20}$ & w & 2247.9 (0.03) & & & w & 2767.4 (0.01) & 2772.8 (0.00) & $\nu_{8}$+$\nu_{13}$ \\
x & 1920.2 (0.01) & 1930.3 (0.01) & 2$\nu_{19}$ & x & 2273.3 (0.01) & 2283.9 (0.01) & $\nu_{14}$+$\nu_{16}$ & x & 2811.3 (0.02) & 2821.7 (0.01) & $\nu_{7}$+$\nu_{13}$ \\
y & 1951.2 (0.04) & 1957.8 (0.03) & $\nu_{18}$+$\nu_{19}$ & y & 3047.7 (0.02) & 3053.4 (0.01) & $\nu_{4}$ & y & 2880.4 (0.01) & 2865.1 (0.00) & $\nu_{6}$+$\nu_{25}$ \\
z & 1983.5 (0.03) & & & z & 3078.0 (0.04) & 3088.1 (0.02) & $\nu_{2}$ & z & 3444.2 (0.01) & 3431.1 (0.01) & $\nu_{6}$+$\nu_{13}$ \\
A & 2243.9 (0.03) & 2227.5 (0.02) & $\nu_{6}$ & A & 3084.1 (0.09) & 3088.1 (0.03) & $\nu_{2}$ & & & & \\
B & 2246.0 (0.02) & 2228.8 (0.02) & $\nu_{5}$ & & & & & & & & \\
C & 3080.5 (0.02) & 3092.6 (0.02) & $\nu_{3}$ & & & & & & & & \\
D & 3083.0 (0.01) & 3093.5 (0.01) & $\nu_{4}$ & & & & & & & & \\
E & 3088.4 (0.05) & 3098.8 (0.03) & $\nu_{1}$ & & & & & & & & 
\enddata
\end{deluxetable*}

%\newpage
\subsection{193 nm photolysis}

\begin{figure}[h!]
    \centering
    \includegraphics[scale =0.35]{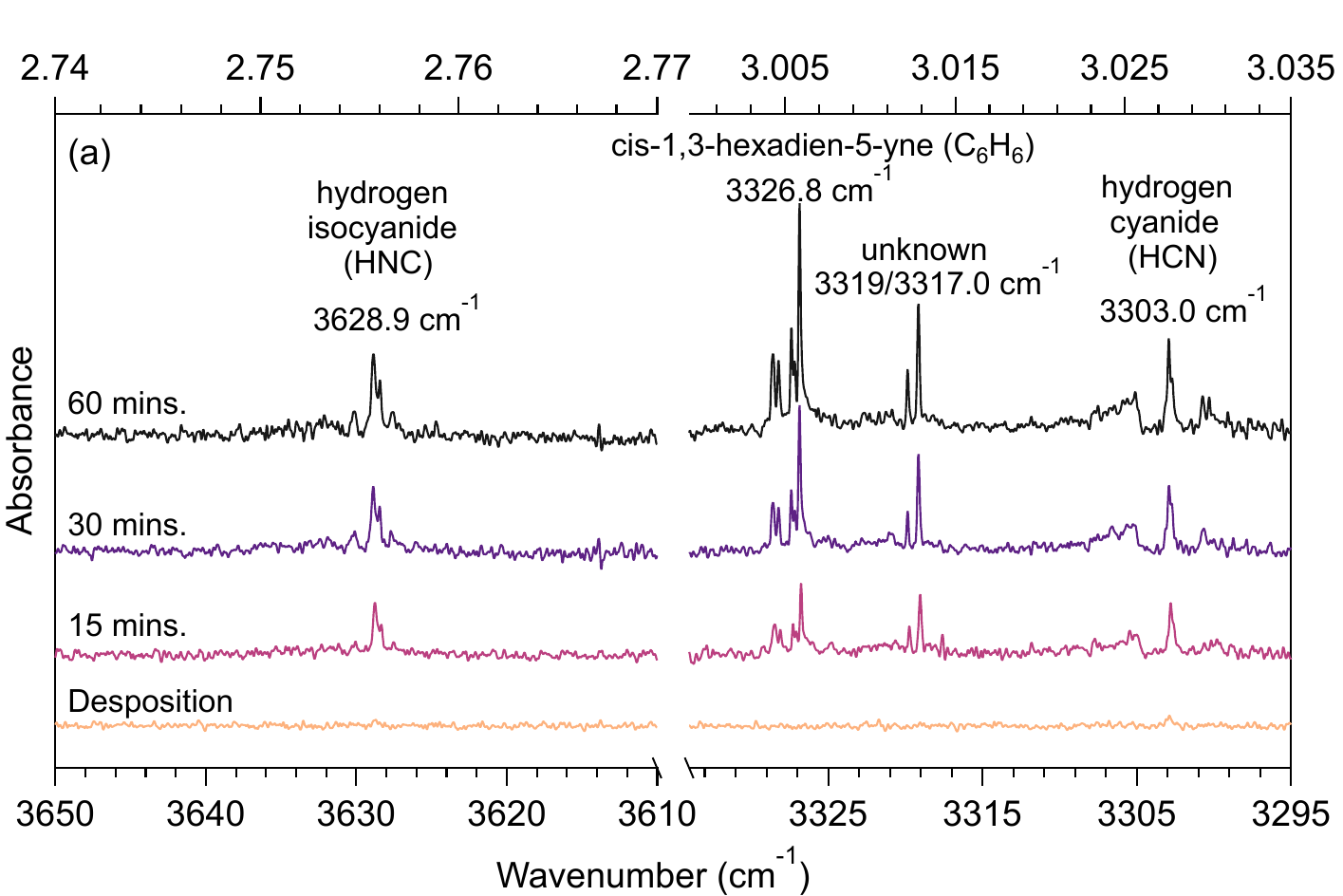}
    \includegraphics[scale =0.35]{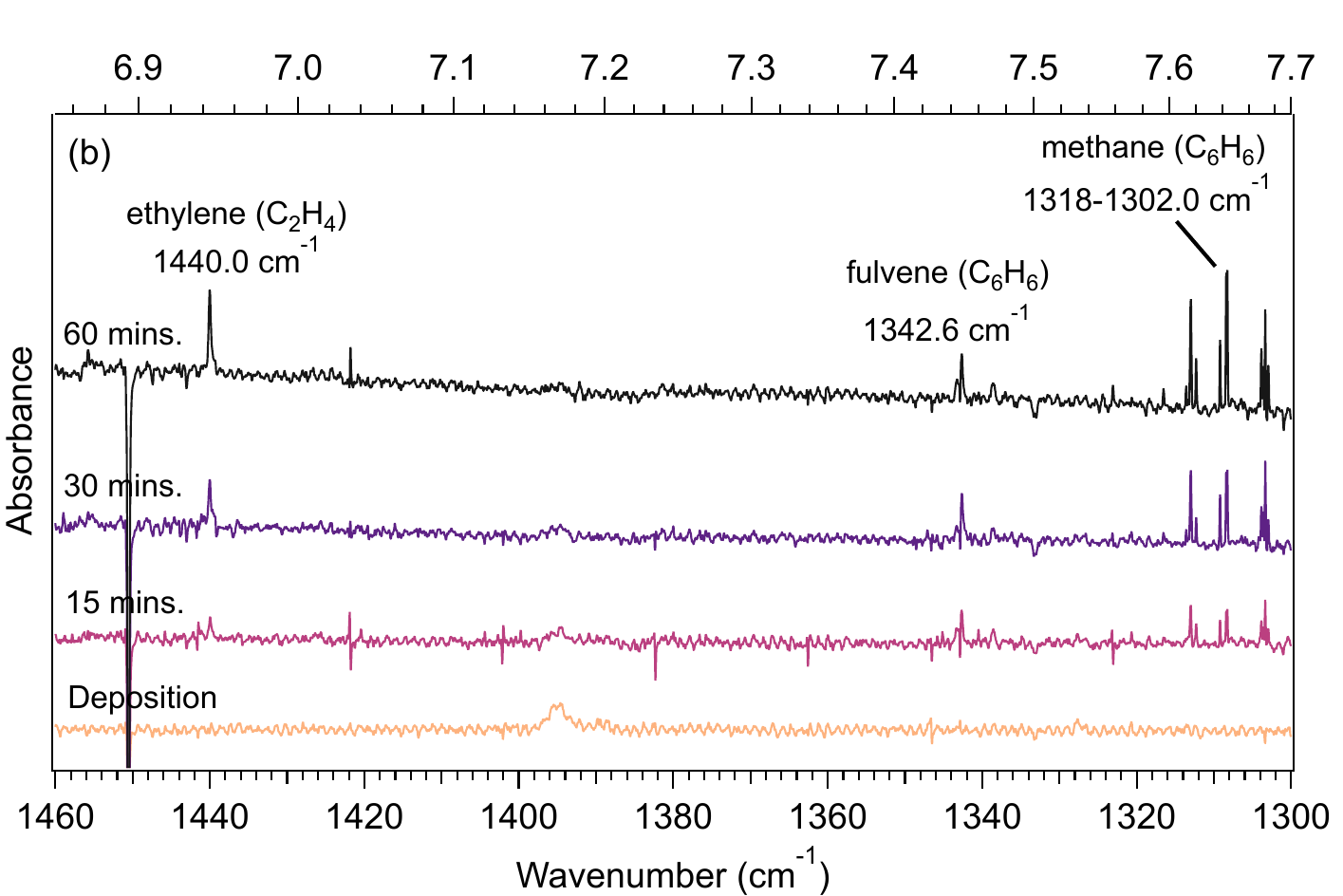}
    \includegraphics[scale =0.35]{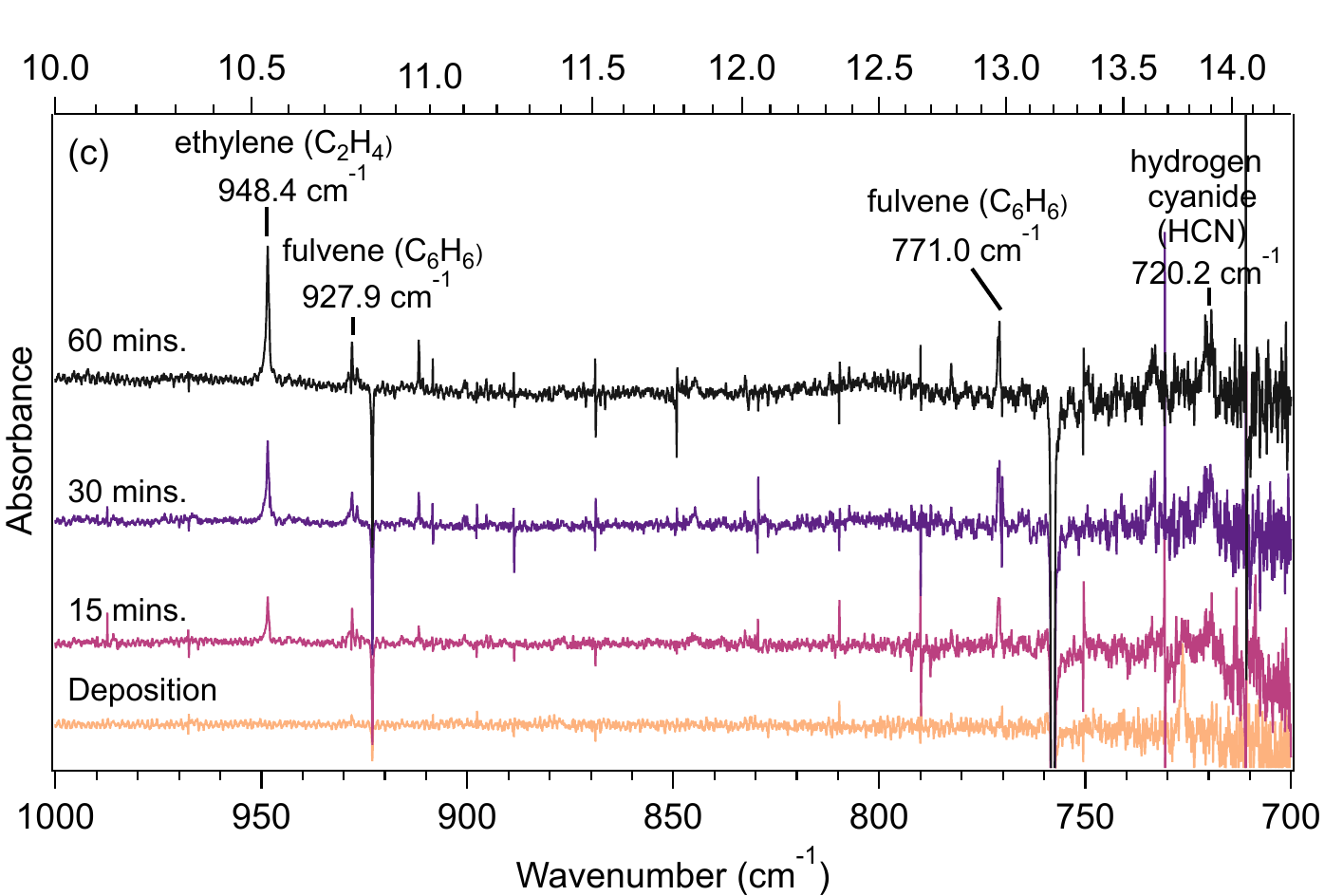}
	\caption{Difference FTIR spectra showing the main photo-products produced from cyanobenzene at deposition and after 15, 30 and 60 minutes of irradiation. Panels are shown in the wavenumber ranges: (a) 3650--3610 and 3335--3295 cm$^{-1}$, (b) 1460--1300 cm$^{-1}$, and (c) 1000--750 cm$^{-1}$.}
	\label{fig:photoproductsspectra}
\end{figure}

All nitrile derivatives were irradiated at 193 nm (6.42~eV) using an ArF excimer laser. The results of the 193 nm photolysis for cyanobenzene can be seen in Fig. \ref{fig:photoproductsspectra}. Difference FTIR spectra are given at different irradiation times, showing the formation of several photoproducts (peaks) and the dissociation of the sample (troughs). Here, it shows the results in the following ranges: 700--1000 cm$^{-1}$, 1300--1460 cm$^{-1}$, 3295--3335 cm$^{-1}$ and 3610--3650 cm$^{-1}$. Spectra were taken at deposition, as well as after 15, 30 and 60 minutes of irradiation time in ascending order.   

Table \ref{tab:minor_products} contains a summary of the observed photoproducts formed from cyanobenzene and the dicyanobenzene isomers. The assignment of each vibrational mode was made based on previous work in p-H$_2$ or other matrices. Theoretical calculations were also used to assist in these assignments.

\begin{figure}[h!]
    \centering
    \includegraphics[scale =0.35]{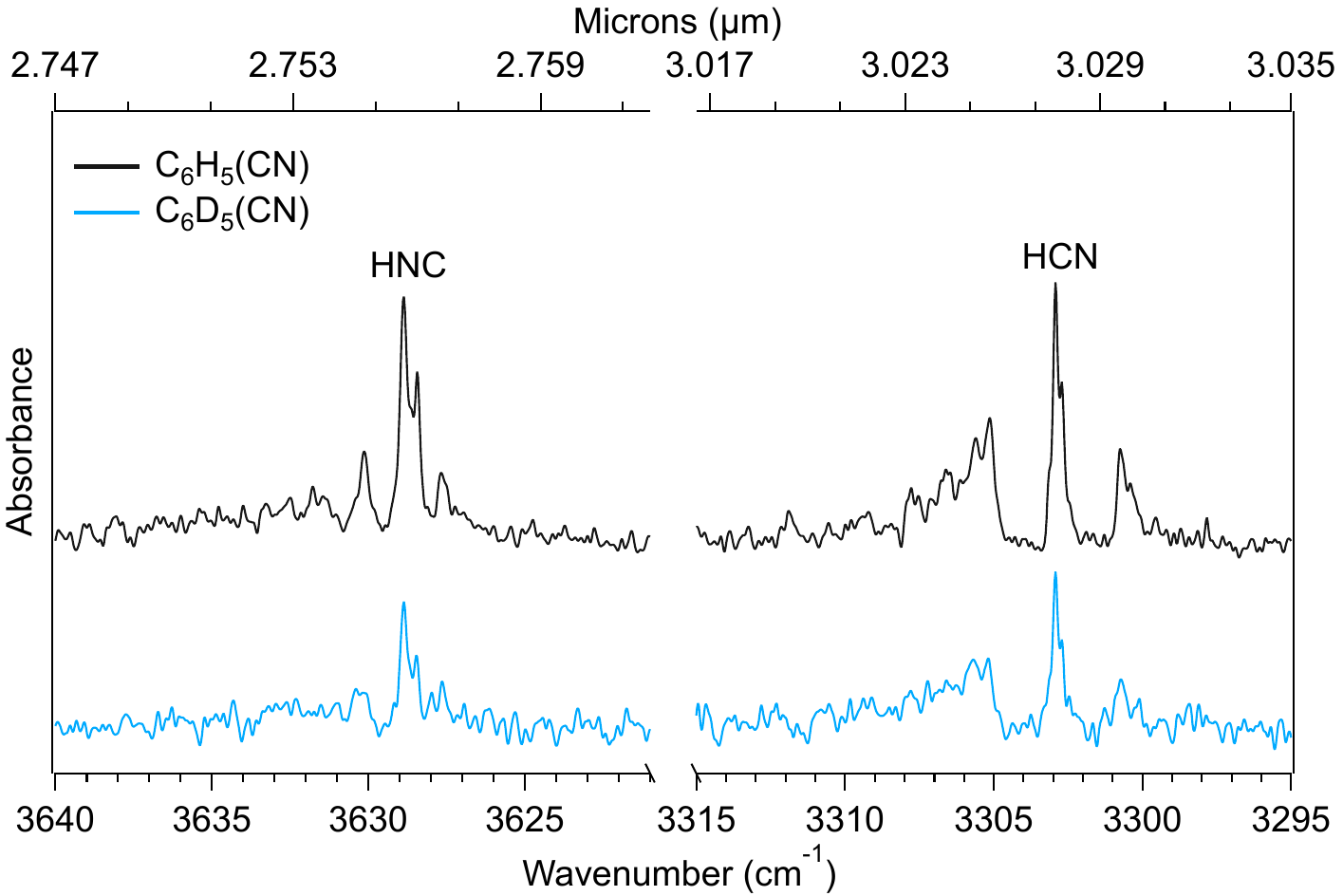}
	\caption{Absorption spectra showing HCN and HNC formed from the photolysis of cyanobenzene (black) and d$_5$-cyanobenzene (blue) in the range of 3295--3315 cm$^{-1}$ and 3620--3640 cm$^{-1}$.}
	\label{fig:HCN_HNC_IR}
\end{figure}

Additional experiments were conducted with deuterated cyanobenzene (d$_5$-cyanobenzene) to elucidate the mechanism through which HCN and HNC form. Fig. \ref{fig:HCN_HNC_IR} shows the spectrum of HCN and HNC from irradiation of both the non-deuterated and deuterated samples.  

\begin{figure}[t!]
    \centering
    \includegraphics[scale =0.35]{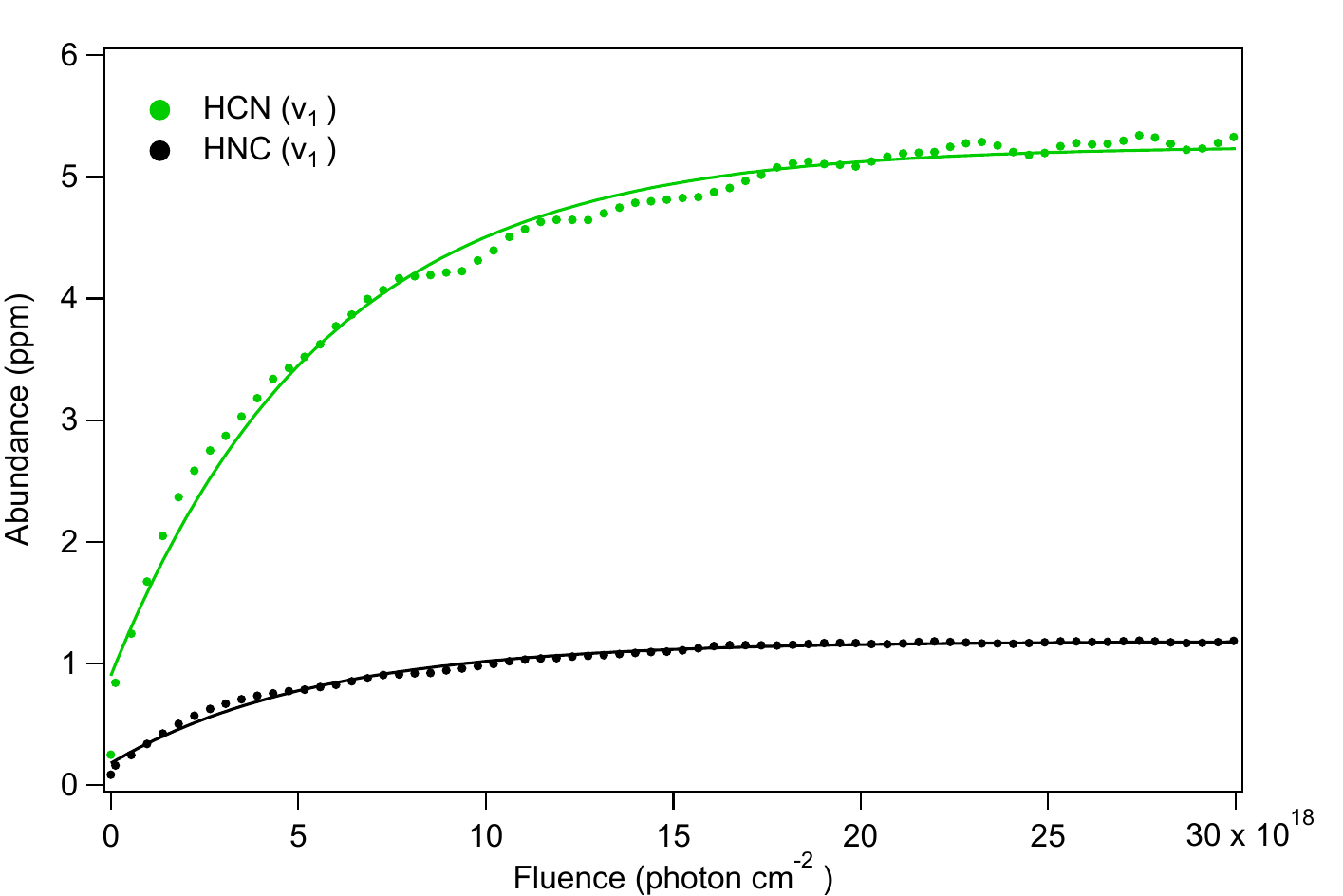}
	\caption{Temporal behaviour of the photoproducts, HCN (green) and HNC (black), formed after irradiating a $\sim$40 ppm cyanobenzene sample at 193 nm for 3 hours.}
	\label{fig:photoproductskinetics}
\end{figure}

Fig. \ref{fig:photoproductskinetics} shows a comparison of the formation of HCN/HNC in both cyanobenzene and 1,3-dicyanobenzene based on their calculated abundances in the p-H$_2$ matrix plotted against fluence. The concentration is calculated in ppm based on the integrated peak intensity of $\nu_{8}$ for HCN (3302.7 cm$^{-1}$) and HNC (3628.4 cm$^{-1}$). The details of these calculations can be found in Appendix \ref{sec:HCN-HNC}. An example of the temporal evolution of the other photoproducts listed in Table \ref{tab:minor_products} is shown in Appendix \ref{sec:time-products}.

\section{Discussion}

\subsection{Spectroscopy of the CN stretches}

The frequency of a vibration is influenced by the medium in which it is observed. For example, the CN stretching mode of cyanobenzene in the gas phase is found at 2229 cm$^{-1}$ and appears as a single band \citep{Rajasekhar2022}. However, within p-H$_2$ matrices, the band is split and shifted to 2238.2 and 2241.9 cm$^{-1}$, which can be seen in Fig. \ref{fig:cn_comparison}. This splitting is also evident in an argon matrix where two peaks are observed at 2237.3 and 2241.4 cm$^{-1}$ \citep{Bernstein1997}. This splitting may be intrinsic to this species and a result of possible tunneling motions in the matrix. Again, this splitting is visible in neon matrices with a larger shift from gas-phase values to 2240.8 and 2245.6 cm$^{-1}$ \citep{Huang2025}. 

Recently, \citet{Esposito2025} have developed a new method for computing the anharmonic vibrational frequencies of cyano-substituted PAHs utilizing a double-hybrid DFT functional with MP2 electron correlation to compute the harmonic frequencies, as well as the cubic and quartic force constants of the B3LYP functional for the anharmonic part of the potential. Using the rDSD-TZ+B3LYP/N07D hybrid method, they computed the CN stretch frequency of cyanobenzene of 2225.8 cm$^{-1}$, which is in excellent agreement with the gas-phase measurements and a difference of 9--13 cm$^{-1}$ from this work in p-H$_2$.  Quantum calculations are considerably intensive for larger molecules, such as PAH nitriles. Experimental spectroscopy thus remains a necessary tool and can be used to benchmark new computational methods.

Given the presence of two CN groups in each dicyanobenzene molecule, two CN stretch fundamentals result from symmetric and anti-symmetric stretching of the CN groups. Indeed, there are two peaks observed in each of the three dicyanobenzene spectra shown in Fig. \ref{fig:cn_comparison}. However, there is a noticeable difference in the frequency and peak separation between isomers. Beginning with 1,2-dicyanobenzene, there are two peaks in the experimental spectrum at 2246.0 and 2243.9~cm$^{-1}$. These bands are assigned to the two CN stretch fundamentals, ID `A' and `B', computed at 2228.8 ($\nu_5$) and 2227.5~cm$^{-1}$ ($\nu_6$), a difference from experiment of 17.2 and 16.4~cm$^{-1}$, respectively. This difference is mostly attributed to matrix effects. 

In 1,3-dicyanobenzene, the peaks are slightly blue-shifted to 2247.3 and 2246.7~cm$^{-1}$, a frequency spacing of only 0.5~cm$^{-1}$. These bands are similarly assigned to the two CN stretch fundamentals, $\nu_5$ and $\nu_6$, computed at 2232.6 and 2231.1~cm$^{-1}$, a difference from experiment of 15.6 and 14.7~cm$^{-1}$, respectively. The intensity of the transitions is flipped in the computed spectrum, indicating a shifting of the peaks, incorrect intensity calculations, or the presence of another transition in the experiment that is not captured in the present computations. 

The 1,4-cyanobenzene spectrum shows a deviation compared to the other two molecules, with a larger frequency spacing of 8.0~cm$^{-1}$ between the peaks at 2250.5 and 2242.6~cm$^{-1}$. This spacing is mostly captured in the computed spectrum, where the assigned transitions `u' and `v' at 2231.9 and 2226.4~cm$^{-1}$ have a spacing of 5.5~cm$^{-1}$ and a difference from experiment of 18.6 and 16.2~cm$^{-1}$, respectively. The underlying normal modes that lead to these transitions differ from those of the other two dicyanobenzene isomers. 1,4-dicyanobenzene is the only $C_{2v}$ isomer, and therefore, only has a single IR-active CN stretch fundamental: the anti-symmetric stretch, $\nu_5$. The symmetric CN stretch ($\nu_6$) leads to no change in the dipole moment. The two experimental bands arise from a complementary resonance-coupled pair of the anti-symmetric CN stretch ($\nu_5$) and the $\nu_{14}$+$\nu_{17}$ combination band. The band at 2242.6~cm$^{-1}$ has a majority contribution from $\nu_5$ while the band at 2250.5~cm$^{-1}$ has a majority contribution from $\nu_{14}$+$\nu_{17}$; although both transitions derive their intensity from the anti-symmetric CN stretch fundamental. All three isomers show a matrix shift of around 15 -- 18~cm$^{-1}$, and future gas-phase experiments would serve to confirm this shift.

\subsection{Photodissociation Kinetics\label{kinetics}}

The electronic spectrum of cyanobenzene has been characterized through both experimental and computational studies \citep{Dixon2015, Rajasekhar2022}. In the range of 111---278 nm (4.5-11 eV), it has been shown to have multiple valence and Rydberg transitions. There are two sharp absorption peaks at 4.49 and 5.53 eV, corresponding to valence $\pi$ $\rightarrow$ $\pi^*$ transitions, as well as a strongly absorbing broad band peaking at 6.62 eV. Analysis of the photoabsorption spectrum of cyanobenzene in the gas phase at 298 K shows it to be strongly absorbing at 193 nm (6.42 eV); with an absorption cross-section of $\sim$$\sigma$ = 3.72 $\times$ 10$^{-17}$ cm$^2$ \citep{Rajasekhar2022}. %$\sigma$ = 3.72 $\times$ 10$^{-17}$ cm$^2$. 

The threshold for dissociation of ground-state cyanobenzene is calculated to be $\sim$5.65 eV from standard enthalpies of formation \citep{Aoyama2005}. Irradiation at 193 nm is sufficiently high for excitation into low-lying triplet states, ultimately leading to the dissociation of the bond between the benzene ring and the CN group. \citep{Rajasekhar2022}.  

Although 213 nm (5.82 eV) is slightly above this threshold, in our experiments no dissociation was observed after 2 hours of irradiation. Furthermore, no new photoproduct signals appear in the IR spectra. To the best of our knowledge, the electronic structure of the dicyanobenzene isomers and their UV absorption have not yet been reported in the literature.  

\subsection{Primary Photoproducts\label{subsec:photoproducts}}

\textbf{Cyanobenzene:}
The primary photoproducts from the photolysis of cyanobenzene at 193 nm are suspected to be the cyano (CN) and phenyl (C$_6$H$_5$) radicals, despite not being directly observed in the IR spectra of our experiments. Often, the photolysis of nitrile derivatives at UV/VUV wavelengths proceeds through the cleavage of the C--C bond connecting the CN group, yielding the CN radical \citep{McElcheran1958,Horwitz1997}. In other cases, H addition to the CN group can occur, yielding HCN as a major photoproduct, as has been observed for vinyl cyanide (CH$_2$CHCN; \citealt{Wilhelm2009,Bird1996}). Cleavage of the CC bonds within the ring is unlikely to occur until the aromaticity is first broken, as discussed for benzene \citep{Kislov2004}. A similar study conducted on the photolysis of cyanobenzene at 193 nm in the gas phase found the major products to be C$_6$H$_5$ and CN \citep{Park1989}. 

CN has been shown to absorb at 2050.8 cm$^{-1}$ in solid H$_2$  \citep{Ruzi2012} and at 2046 cm$^{-1}$ in an argon matrix \citep{Milligan1967} but is not seen in our experiments. However, the presence of CN is evident through the formation of both HCN and HNC, which are observed at 3302.9 cm$^{-1}$ and 3628.8 cm$^{-1}$, respectively, in Fig. \ref{fig:HCN_HNC_IR}. Both molecules are sufficiently small to allow some amount of rotation within the matrix, permitting observation of ro-vibrational spectral structure. Our results suggest that CN reacts with the solid para-H$_2$ to give HCN and HNC. A similar reaction pathway was previously investigated for CN produced from the photolysis of cyanogen (C$_2$N$_2$; \citealt{Borget2017}), when reacting with solid H$_2$. Reactions involving the deuterated methyl radical (CD$_3$; \citealt{Momose1998}) and the hydroxyl radical (OH; \citealt{Oba2012}) inside a para-H$_2$ matrix have been suggested to occur via quantum tunneling, which may likewise explain the formation of HCN and HNC observed here. 

In order to elucidate the mechanism of HCN/HNC formation, d$_5$-cyanobenzene (C$_6$D$_6$CN) was irradiated at 193 nm. Once again, the same HCN/HNC vibrational modes appear, which are visible in Fig. \ref{fig:HCN_HNC_IR}. None of the corresponding peaks for DCN or DNC were seen in the spectra compared to previous assignments in solid hydrogen \citep{Borget2017}, implying that HCN/HNC is formed by reaction with the matrix. 

One alternate formation pathway could be that a C-H bond on the ring is photolyzed and the liberated H atom subsequently reacts with CN. Though H loss is not expected to be a dominant channel in the 193 nm photolysis of cyanobenzene itself, H loss from phenyl is expected to be a major channel \citep{Tseng2004,Mebel2012}. However, the rapid onset of HCN (and HNC) in our experiments is typical of primary photoproducts, thus the formation from a mechanism involving phenyl appears less likely. This pathway is also possible for (C$_6$D$_6$), although due to the lower zero-point vibrational energy causing the C-D bond to be more stable, C-D bond breaking is less favorable. A second alternative mechanism is H migration via H-atom tunneling from the aromatic ring to either the C or N atoms prior to dissociation. The heavier D atom has much less efficient tunneling probability, lowering the likelihood of a D migration pathway forming DCN/DNC. Further explorations of the differences in the photodynamics of normal and deuterated aromatics would aid in confirming the origin of the HCN/HNC photoproducts.

Based on assignments in an argon matrix \citep{Friderichsen2001} the phenyl radical C$_6$H$_5$ strongly absorbs at 703--706 cm\textsuperscript{-1} and at 3071--3073 cm\textsuperscript{-1}, neither of which was seen at any time throughout irradiation. Possibly due to reactions with the matrix or liberated H atoms to give benzene, which is subsequently photolyzed at 193 nm. Weak signals associated with benzene's $\nu_{14}$ (1038 cm\textsuperscript{-1}) and $\nu_{13}$ (1482.9cm\textsuperscript{-1}) vibrational mode are potentially present at low signal-to-noise. Alternatively, as mentioned above, phenyl itself may be dissociated through absorption at 193 nm. Previous experiments have shown that phenyl dissociates at 193 nm through different channels, producing either o-C$_6$H$_4$/l-C$_6$H$_4$ + H or C$_4$H$_3$ (butatrienyl radical) + C$_2$H$_2$ \citep{Tseng2004,Negru2010, Mebel2012, Song2012}. We do not see these products in any of our spectra, suggesting it is more likely that the former mechanism is occurring. 

%Furthermore, the cis-isomer of 1,3-hexadien-5-yne is observed to form supporting the suspected formation of a C$_6$H$_6$ species. However, the trans-isomer is not present in the case of cyanobenzene. 

\textbf{Dicyanobenzene isomers:}
The photoproducts formed during the 193 nm irradiation are the same for each of the dicyanobenzene isomers and mirror those of cyanobenzene closely. The vibrational modes for HCN and HNC appear earlier than any other photoproduct according to kinetic analysis. Again, proceeding through the dissociation of the bond between the CN group and the aromatic ring. 

The fate of the aromatic radical species formed from this is not fully understood. It is thought that cyanobenzene or a related species is first formed, which then undergoes a secondary photolysis process. Consecutive CN cleavages and H reactions then yield photoproducts that overlap with those found for cyanobenzene.

The concentration (ppm) of each isomer has been estimated for all photolysis experiments by integrating the modes $\nu_{1}$ for HCN and $\nu_{1}$ for HNC, respectively. The parameters used for these calculations are given in \ref{tab:ppm_parameters}, resulting in HNC/HCN ratios in the range of $\sim$0.22--0.35. As they are not at thermal equilibrium, it is suspected that this ratio results from kinetic control governed by the rates of each reaction.

\begin{deluxetable}{ccccc}[hbt!]
\tablecaption{Summary of the assigned vibrational modes of photoproducts observed during the irradiation of all nitrile derivatives in cm$^{-1}$. As well as the associated vibrational modes for each transition and their previous assignments in p-H$_2$. $\ast$ denotes that it was only observed in cyanobenzene and not the dicyanobenzene isomers.}
\label{tab:minor_products}
\tablehead{    
\colhead{Molecule}&
\colhead{p-H$_2$}& 
\colhead{Assignment} & 
\colhead{Mode} & 
\colhead{R}}
\startdata   
hydrogen cyanide & 3302.7 & $\nu$(C-H) & $\nu_{1}$ & [a]\\
(HCN) & $\cdots$ & $\nu$(C-N) & $\nu_{2}$ & \\
 & 720.2 & $\delta$(H-C-N) & $\nu_{3}$ & \\
 & 720.2 & $\delta$(H-C-N)  & $\nu_{4}$ & \\
\hline
hydrogen isocyanide  & 3628.4 & $\nu$(N-H) & $\nu_{1}$ & [a] \\
(HNC) & 2025.1 & $\nu$(C-N) & $\nu_{2}$ &  \\
 & $\cdots$ & $\delta$(H-N-C) & $\nu_{3}$ & \\
 & $\cdots$ & $\delta$(H-N-C) & $\nu_{4}$ &  \\
 \hline
cis-1,3-hexadien-5-yne & 3327.4 & $\nu$(C-H) & $\cdots$ & [b] \\
(C$_6$H$_6$) & 3326.8 & $\nu$(C-H)  & $\cdots$ & \\
\hline
fulvene $\ast$  & 1342.6 & $\delta$(C-H) & $\nu_{7}$ & [b] \\
(C$_6$H$_6$) & 927.9 & $\omega$(CH$_2$) & $\nu_{16}$ & \\
 & 895.3 & $\nu$(C-C-C) & $\nu_{10}$ & \\
 & 771 & $\omega$(C-H) & $\nu_{18}$ & \\
\hline
ethylene & 2984.2 & $\nu$(CH$_2$) & $\nu_{11}$ & [c] \\
(C$_2$H$_4$) & 1887.1 & $\cdots$ & $\nu_{7}$ + $\nu_{8}$ & \\
 & 1440 & $\delta$(CH$_2$) & $\nu_{12}$ & \\
 & 948.4 & $\omega$(CH$_2$) & $\nu_{7}$ & \\
\hline
methane & 3026 & $\nu$(CH$_4$) & $\nu_{3}$ & [d] \\
(CH$_4$) & 3025.2 & $\nu$(CH$_4$) & $\nu_{3}$ & \\
 & 1309.2 & $\delta$(CH$_4$) & $\nu_{4}$ & \\
 & 1308.4 & $\delta$(CH$_4$) & $\nu_{4}$ & \\
 & 1308.2 & $\delta$(CH$_4$) & $\nu_{4}$ & \\
 \hline
unknown & 3319.9 & $\cdots$ & $\cdots$ &  \\
 & 3319.2 & $\cdots$ & $\cdots$ &  \\
 & 2270.9 & $\cdots$ & $\cdots$ & \\
 \hline
\enddata
\tablenotetext{}{References: \citet{Borget2017} [b] \citet{Toh2015}, [c] \citet{Pinelo2018}, [d] \citet{Momose1997} }
\end{deluxetable}

\subsubsection{Secondary/minor photoproducts}

The presence of the secondary photoproducts fulvene and cis-1,3-hexadien-5-yne also supports the formation of benzene, as both are products of the 193 nm photolysis of benzene. Fulvene is seen to form after prolonged irradiation, as evidenced by its most intensely absorbing mode, $\nu_{14}$ at 771 cm\textsuperscript{-1}. However, none of the most intensely absorbing peaks for benzvalene or Dewar benzene are present, which is the case for pure benzene photolysis at 193 nm \citep{Toh2015}. 

One difference between the 193 nm photolysis of benzene and cyanobenzene is the formation of ethylene (C$_2$H$_4$). This assignment is based on previous characterisation of ethylene in p-H$_2$ \citep{Pinelo2018}, where the following modes are seen in our spectra: $\nu_{7}$ (948.4 cm$^{-1}$), $\nu_{12}$ (1440 cm$^{-1}$), $\nu_{7}$ + $\nu_{8}$ (1887.1 cm$^{-1}$) and $\nu_{11}$ (2984.2 cm$^{-1}$). This suggests there may be another dissociation pathway involving a ring-opening and fragmentation of the carbon chain. 

The accompanying fragment from this is possibly associated with the currently unknown peaks at 3319.9 cm$^{-1}$ and 3319.2 cm$^{-1}$. This species is formed in the photolysis of cyanobenzene, 1,2-dicyanobenzene, 1,3-dicyanobenzene and 1,4-dicyanobenzene. Kinetic analysis suggests that it is related to the peak at 2270.9 cm$^{-1}$. The positioning of these frequencies implies that it could be an alkyne nitrile, most likely a cyanopolyyne such as cyanoacetylene (HC$_3$N) due to previous assignments in other matrices. In an argon matrix,  HC$_3$N absorbs at 3316 cm$^{-1}$, 2269 cm$^{-1}$, 2076 cm$^{-1}$ and a final peak at 667 cm$^{-1}$ \citep{Borget2001}. To confirm the assignment, we ran an additional experiment using a Zinc Selenide (ZnSe) window, enabling observation of modes at longer wavelengths. Despite this, we did not observe the $\nu_5$ band nor its first overtone (1318 cm$^{-1}$ in Ar); however, a similar non-detection of these modes was reported by \citet{Coupeaud2006} for photolytically produced HC$_3$N. These authors attributed the non-detection to the formation of HC$_3$N in a tight matrix site that would inhibit bending motion.  

Finally, one curious photoproduct formed at later stages of irradiation is methane (CH$_4$). The mechanism of its formation is still not fully understood, particularly due to it also appearing in the photolysis of d$_5$-cyanobenzene. All deuterated forms of methane have been characterized in para-H$_2$ \citep{Hoshina1999}, none of which are apparent in our study. Therefore, we anticipate it may form through reactions with the para-H$_2$ matrix. 

\subsection{Astrophysical Implications}

It is now well-established that aromatic molecules are abundant in dense molecular clouds. Benzene and its nitrile counterpart, cyanobenzene, represent the smallest aromatics detected to date and likely play a key role in the formation of PAHs. It is thought that PAHs produced in the circumstellar envelope may be inherited into these colder environments. However, for this to be the case, they must survive harsh radiation fields on their journey. Even within these cold clouds, they are subjected to VUV radiation. 
Suspected PAH emissions via the AIB bands are found throughout the ISM at different stages of the stellar lifecycle. Absorption of FUV photons leads to vibrational excitation of PAHs over their many vibrational modes via intramolecular vibrational energy redistribution. Subsequent relaxation via emission of IR photons from the many possible fundamental, combination and overtone transitions produces mid-, near, and far-IR light, often appearing as a continuum in the regions of 1--5 $\mu$m with more intense signals at 5--15 $\mu$m. PAH models predict that the weak spectral features in the 1--5 $\mu$m range contain fundamental information about the PAH population.

With the emergence of space-based telescopes such as the JWST, high-resolution IR data of astrophysical environments are becoming more readily available. JWST can provide sensitive measurements via its NIRspec instrument in the region of 0.6--5.3 $\mu$m, which is a key region for observing PAH features, including CN-PAH fundamental stretching modes \citep{Allamandola2021,Boersma_2023}. Both PAH nitriles and deuterated PAHs have spectroscopic signatures in the 4.3--4.8 $\mu$m region.  As shown in this work, the wavelength and structure of aromatic CN stretching modes depend on the parent species and the matrix environment. Our work provides critical constraints on the CN stretching modes of single-ring aromatic nitriles that may aid analysis of JWST data. Benchmarking theory against experimental data and understanding matrix shifts in experiments will be critical in establishing a large body of spectroscopic data on aromatic molecules and their derivatives.  This work will be extended to aromatic nitriles containing multiple rings. Further experimental and theoretical spectra are required to compare with observational IR data and ultimately identify trends in PAH chemistry during the different stages of star formation.

Photolysis of CN derivatives has been explored experimentally for smaller saturated and unsaturated species, but less work exists on the photochemistry of aromatic nitriles, particularly PAHs. Our work may be used to better constrain the major photolytic pathways involving aromatic nitriles, including astrochemical networks and chemical models of Titan's atmosphere \citep{LOISON201955}.  Nitrile derivatives are becoming more widely used as observable proxies to estimate the abundances of their parent hydrocarbons. Thus, constraining their destruction pathways and their photoproducts is important to developing a complete picture of the lifecycle of aromatic nitriles in the ISM.

\section{Conclusion}

Several aromatic nitriles have been detected in the cold ISM in the past five years via their rotational fingerprints. In tandem, JWST has returned new, highly sensitive observations of PAH infrared emission towards a range of environments, particularly photodissociation regions that are exposed to intense UV fields. To aid the interpretation of these observations, we have measured the vibrational spectroscopy and photochemistry of singly and doubly cyano-substituted benzene in solid para-hydrogen matrices. These experiments enable us to determine the major products produced during the 193 nm photodissociation of the aromatic nitriles. In all cases, we observe HCN and HNC, which are likely produced by hydrogen abstraction from para-H$_2$ by the CN radical. We conclude that the major photodissociation channel is the formation of the phenyl/cyanophenyl and CN radicals, preserving the aromatic backbone. 

\section{Acknowledgements}

I.R.C. and T.M. acknowledge support from the University of British Columbia and the Natural Sciences and Engineering Research Council of Canada (RGPIN-2022-04684). I.R.C. and T.H.S. acknowledge the support of the Canadian Space Agency through grant 24AO3UBC14 and this work was in part supported through the computational resources and services provided by Advanced Research Computing at the University of British Columbia. V.J.E. acknowledges Chapman University for support. 

\bibliography{PaperRef1.bib, bibliography.bib}
\bibliographystyle{aasjournal}

\newpage

\appendix

\renewcommand\thefigure{\thesection\arabic{figure}}   
\renewcommand\thetable{\thesection\arabic{table}}    

\setcounter{figure}{0}    
\setcounter{table}{0} 

%\section{Deposition spectra}\label{sec:dep-spectra}

\section{Example deposition spectra}\label{sec:dep-spectra}

Sample spectra taken after deposition (prior to photolysis) are shown for cyanobenzene,  d$_5$-cyanobenzene, 1,2-dicyanobenzene, 1,3-dicyanobenzene and 1,4-dicyanobenzene in p-H$_2$ matrices. In the case of cyanobenzene and d$_5$-cyanobenzene, samples were deposited at $\sim$40 ppm (See Appendix C). However, as the dicyanobenzene isomers are solids, accurate deposition concentrations are more difficult to deduce. For these, deposition parameters are optimized to reduce clustering and yield isolated sample molecules. Each sample is compared with theoretical predictions shown in red (see main text for description). Some wavenumber chunks are excluded where no peaks are observed in the experimental spectra. The raw IR data is available on Zenodo
\citep{zenodo_dataset}.

\begin{figure}[hbt!]
\centering
\includegraphics[width=0.8\textwidth]{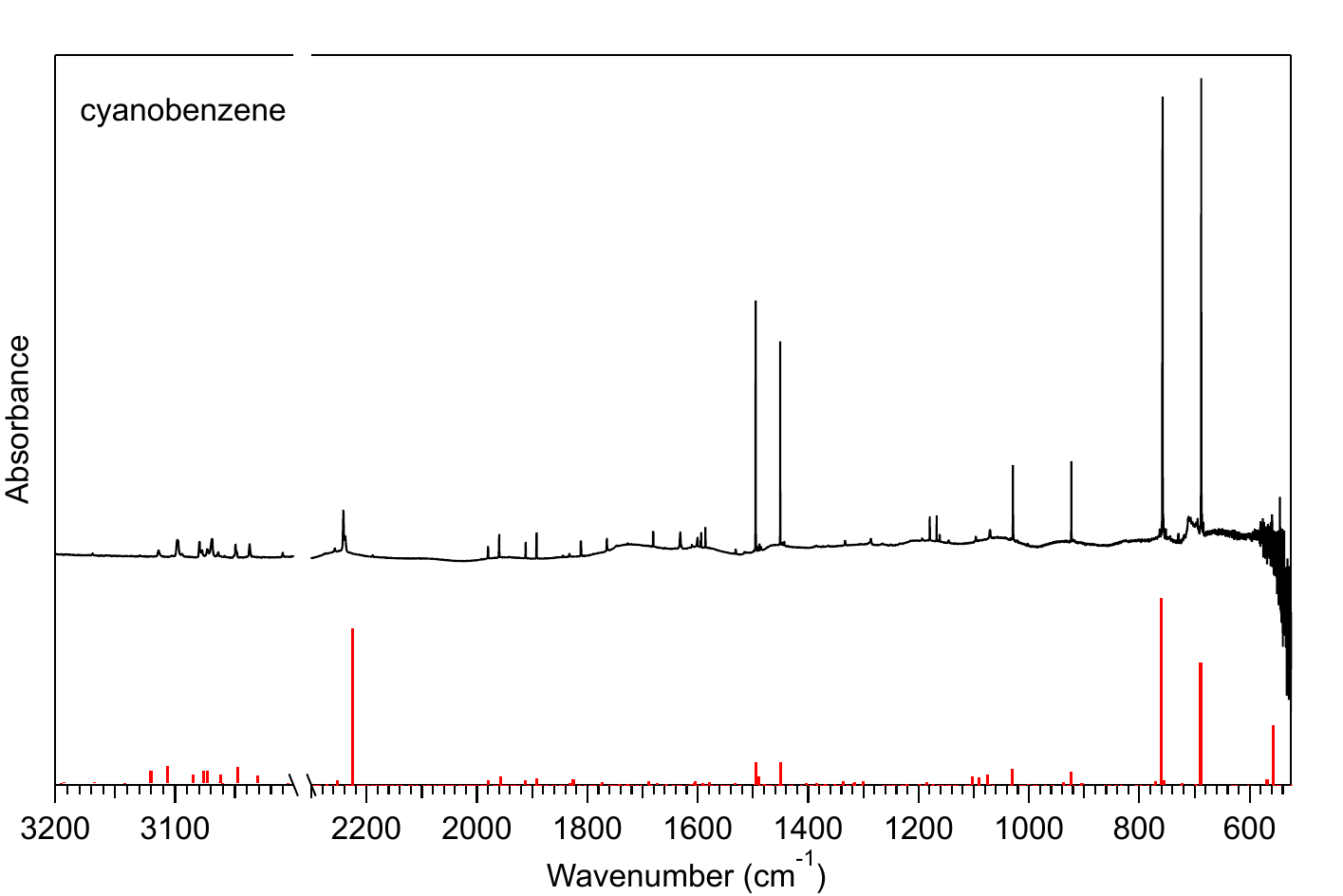}
\caption{Deposition spectrum of cyanobenzene in the range of \textbf{525--2300 cm$^{-1}$} and 3000--3200 cm$^{-1}$ isolated in a para-H$_2$ matrix.}
\label{fig:benzonitriledep}
\end{figure}

\begin{figure}[hbt!]
\centering
\includegraphics[width=0.8\textwidth]{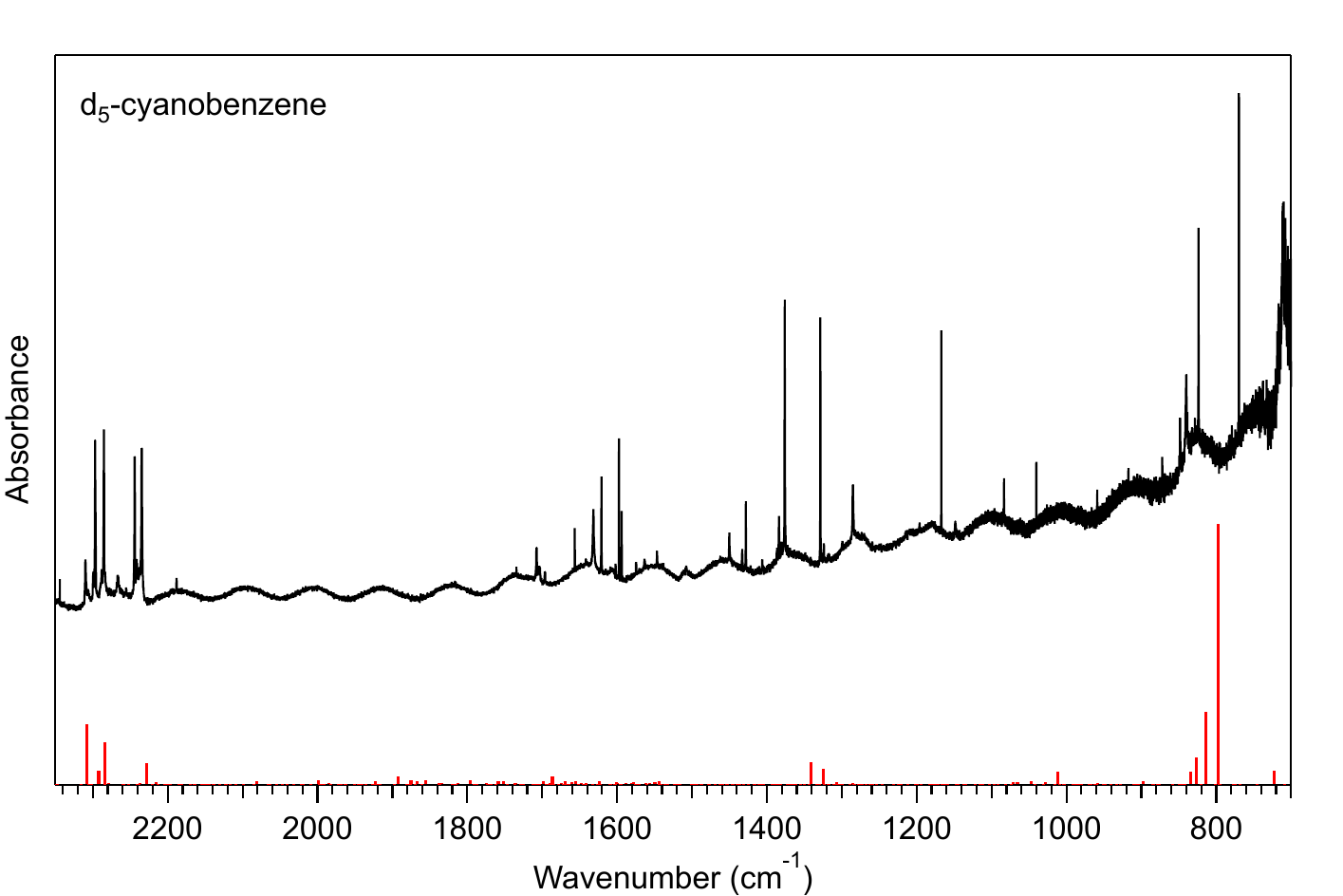}
\caption{Deposition spectrum of d$_{5}$-cyanobenzene in the range of 700--2350 cm$^{-1}$ isolated in a para-H$_2$ matrix.}
\label{fig:perdeuterateddep}
\end{figure}

\begin{figure}[hbt!]
\centering
\includegraphics[width=0.8\textwidth]{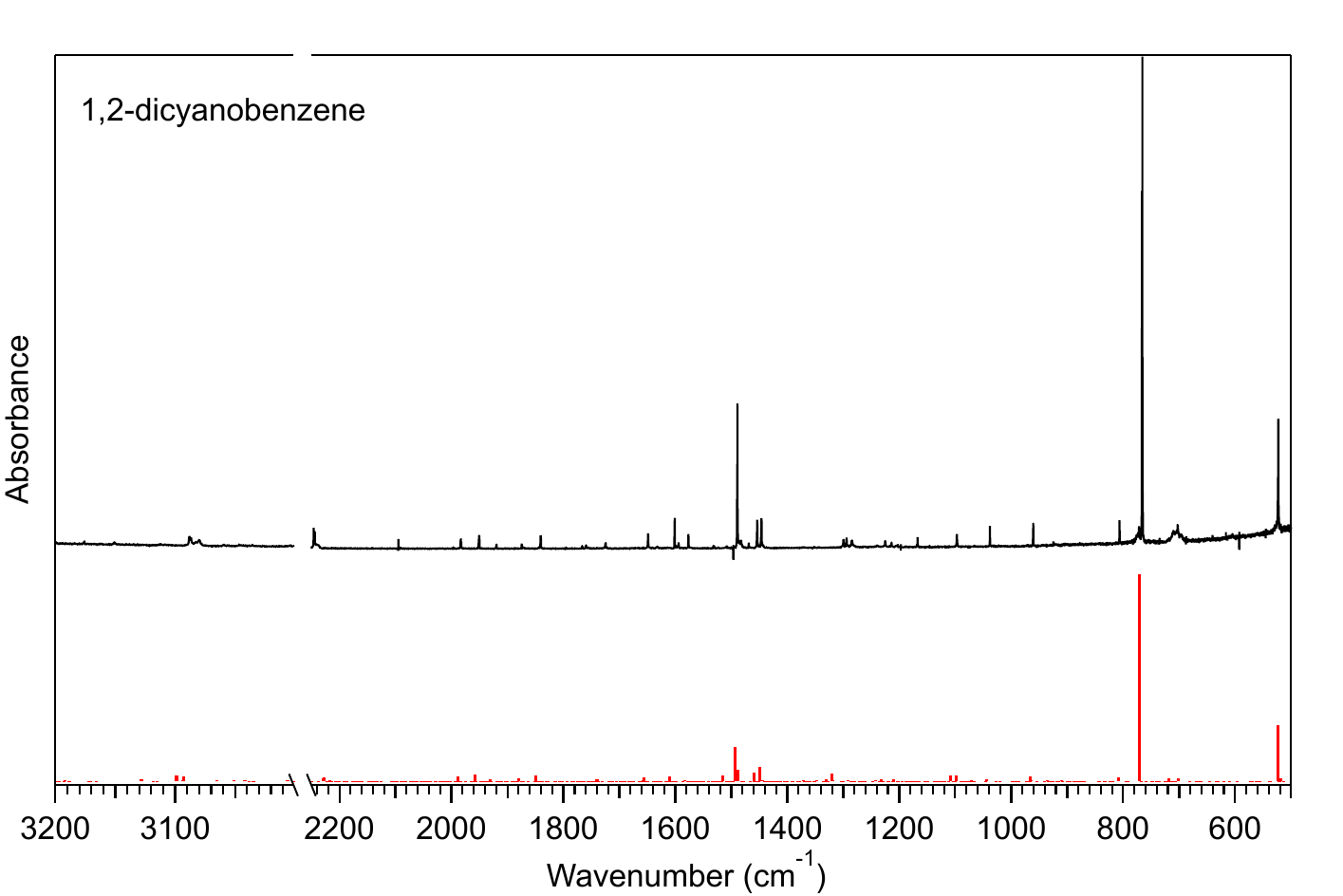}
\caption{Deposition spectra of 1,2-dicyanobenzene in the range of 500--2250 cm$^{-1}$ and 3000--3200 cm$^{-1}$ isolated in a para-H$_2$ matrix.}
\label{fig:12dicyanodep}
\end{figure}

\begin{figure}[hbt!]
\centering
\includegraphics[width=0.8\textwidth]{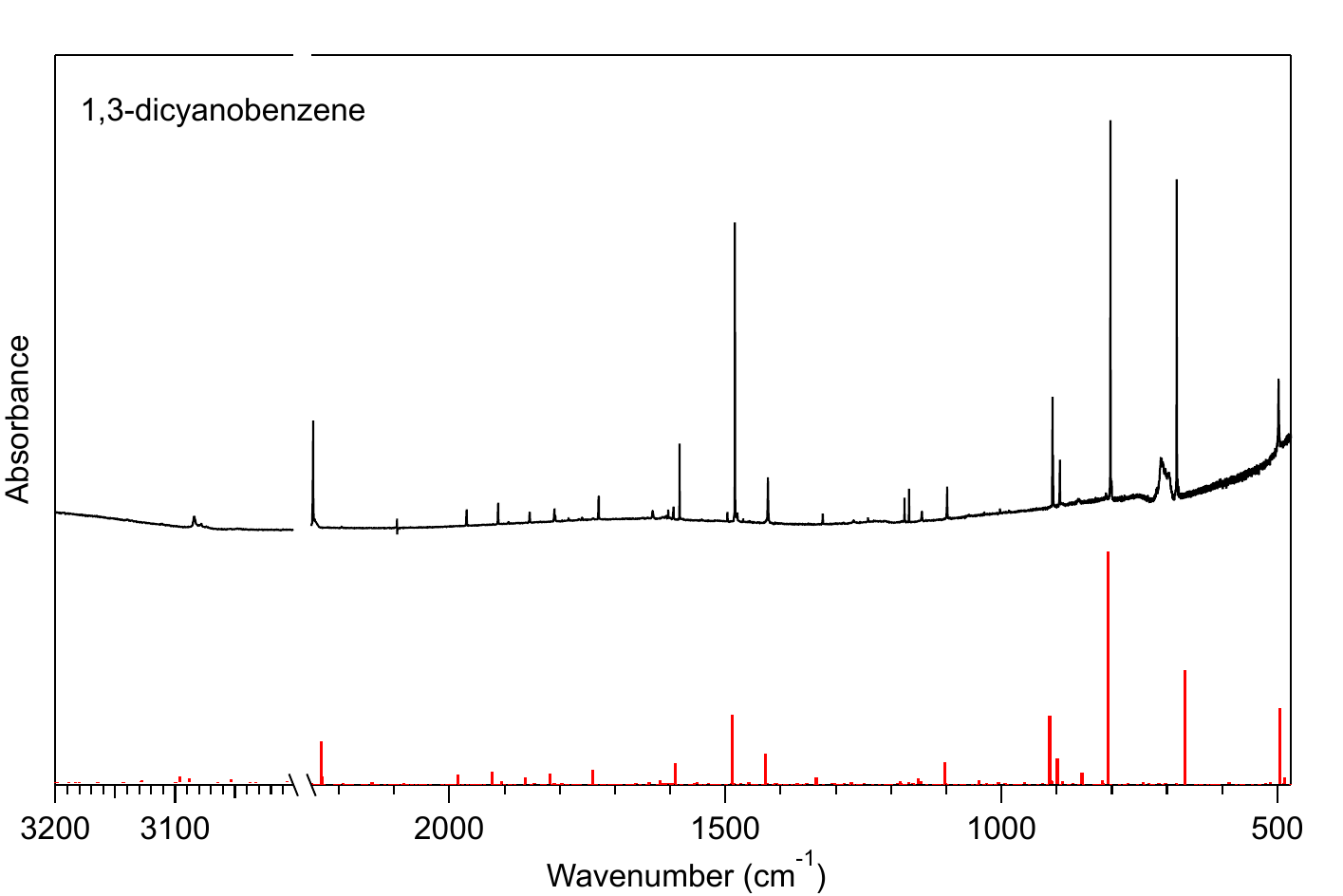}
\caption{Deposition spectrum of 1,3-dicyanobenzene in the range of \textbf{475--2250 cm$^{-1}$} and 3000--3200 cm$^{-1}$ isolated in a para-H$_2$ matrix.}
\label{fig:13dicyanodep}
\end{figure}

\begin{figure}[hbt!]
\centering
\includegraphics[width=0.8\textwidth]{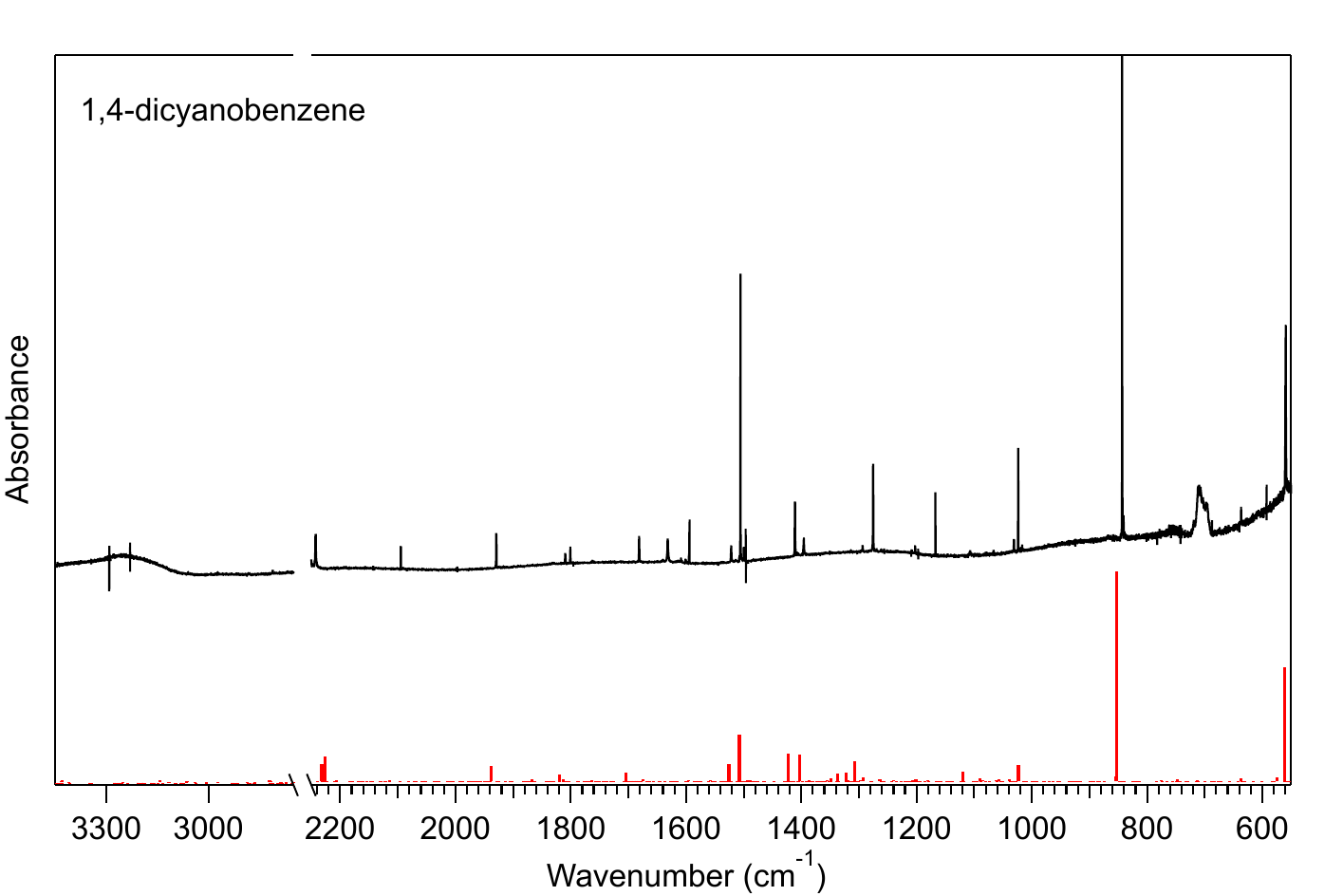}
\caption{Deposition spectrum of 1,4-dicyanobenzene in the range of 550--2250 cm$^{-1}$ and 2750--3450 cm$^{-1}$ isolated in a para-H$_2$ matrix.}
\label{fig:14dicyanodep}
\end{figure}

\newpage
\section{Observed IR transitions for  d$_5$-cyanobenzene}\label{sec:dep-spectra}

Table \ref{table:d_cyano_dep} lists the observed transitions for d$_5$-cyanobenzene compared with the theoretical predictions (see main text for description). The relative intensities of each transition are given in brackets, as well as tentative assignments of their modes.  

\begin{deluxetable}{cccc}[t!]
\tablecaption{Measured IR transitions in cm$^{-1}$, anharmonic computed vibrational transitions in cm$^{-1}$, the corresponding vibrational mode assignments and relative intensities for d$_{5}$-cyanobenzene in solid para-H$_{2}$. Anharmonic data are computed at the rDSD/junTZ+B3LYP/N07D level of theory. }
\label{table:d_cyano_dep}
\tablehead{ 
\multicolumn{4}{c}{\centering d$_{5}$-cyanobenzene} \\ 
\colhead{ID}&
\colhead{p-H$_2$ (I)}&
\colhead{Calc. (I)}& 
\colhead{Mode}}
\startdata
a & 769.6 (0.36) & 797.7 (1.00) & $\nu_{20}$\\
b & 823.6 (0.41) & 814.3 (0.28) & $\nu_{19}$ \\ 
c & 840.0 (0.84) & 826.7 (0.10) & $\nu_{18}$ \\
d & 848.2 (0.05) & 833.5 (0.05) & $\nu_{17}$ \\
e & 872.0 (0.05) & 897.1 (0.01) & 2$\nu_{29}$ \\
f & 958.9 (0.03) & 958.8 (0.01) & $\nu_{25}$+$\nu_{30}$ \\
g & 1040.2 (0.06) & 1011.5 (0.05) & $\nu_{15}$+$\nu_{33}$ \\
h & 1083.3 (0.10) & 1047.4 (0.01) & $\nu_{26}$+$\nu_{28}$ \\
i & 1148.3 (0.06) & & \\
j & 1285.0 (0.51) & 1307.2 (0.01) & $\nu_{23}$+$\nu_{24}$ \\
k & 1328.6 (0.52) & 1324.5 (0.06) & $\nu_{11}$ \\
l & 1376.0 (0.80) & 1341.3 (0.09) & $\nu_{10}$ \\
m & 1383.8 (0.12) & & \\
n & 1428.0 (0.13) & & \\
o & 1432.8 (0.06) & & \\
p & 1450.0 (0.14) & & \\
q & 1546.5 (0.06) & & \\
r & 1563.3 (0.07) & 1549.4 (0.01) & $\nu_{18}$+$\nu_{22}$ \\
s & 1597.4 (0.16) & 1600.5 (0.01) & 2$\nu_{20}$ \\
t & 1620.8 (0.20) & 1623.8 (0.01) & 2$\nu_{19}$ \\
u & 1656.5 (0.05) & 1655.6 (0.01) & 2$\nu_{18}$ \\
v & 1696.3 (0.03) & 1686.3 (0.03) & $\nu_{16}$+$\nu_{17}$ \\
w & 1703.5 (0.13) & & \\
x & 1707.5 (0.24) & &  \\
y & 2234.6 (1.00) & 2227.5 (0.08) & $\nu_{6}$ \\
z & 2237.7 (0.28) & & \\
A & 2241.3 (0.38) & & \\
B & 2244.1 (0.62) & 2236.5 (0.01) & $\nu_{7}$+$\nu_{24}$ \\
C & 2266.5 (0.22) & & \\
D & 2285.2 (0.67) & 2283.9 (0.16) & $\nu_{2}$/$\nu_{4}$ \\
E & 2288.1 (0.12) & & \\
F & 2296.8 (0.81) & 2307.7 (0.23) & $\nu_{4}$/$\nu_{2}$ \\
G & 2299.3 (0.12) & & \\
  & 2309.8 (0.30) & & 
\enddata
\end{deluxetable}

\section{Relative concentrations of HCN and HNC}\label{sec:HCN-HNC}

The integrated IR absorbances of the $\nu$(C-H) mode of HCN and the $\nu$(N-H) mode of HNC were used to estimate their concentration (in ppm) in the para-H$_2$ matrix. 

The concentration of each species, X, are estimated using Beer's law and the thickness of the p-H$_2$ crystal, following equation \ref{eq:ppm}:

\begin{equation}\label{eq:ppm}
    [X] = \frac{2.303 \int Abs(\tilde{\nu}) \, d\tilde{\nu}}{1.06 \,\varepsilon({\textrm{cm mol}^{-1}})\,d(\textrm{cm})}\frac{N_\textrm{A}}{N_{\textrm{pH}_2}}(1 \times 10^6)
\end{equation}

Where Abs is the absorbance at a particular wavenumber ($\tilde{\nu}$), $\varepsilon$ is the integrated absorption coefficient, and $d$ is IR path length, corresponding to the crystal thickness of the p-H$_2$, which is determined through integration
of the parahydrogen Q1(0) + S0(0) transition (4495--4520 cm$^{-1}$). $N_A$ is Avogadro's number (6.022 $\times$ 10$^{23}$ mol$^{-1}$), N$_{pH_2}$ is the number density of solid parahydrogen (2.60 $\times$ 10$^{22}$ cm$^{-3}$), 1.06 corrects for the index of refraction of solid parahydrogen and 2.203 converts the integrated absorbance to optical depth \citep{Tam2001}.

%The coefficient 23.16 cm$^3$ mol$^{-1}$ is the molar volume of p-H$_2$.

\begin{deluxetable}{ccccc}[h!]
\tablecaption{Integrated absorption coefficient and integration ranges used to determine concentrations of HCN and HNC.}
\label{tab:ppm_parameters}
\tablehead{    
\colhead{Molecule}&
\colhead{Mode} & 
\colhead{$\varepsilon$(km \, mol$^{-1}$)} & 
\colhead{Integration range (cm$^{-1}$)}& 
\colhead{Ref.}}
\startdata   
HCN &  $\nu_{1}$ &  60.83 & 3302.1--3303.2  & [a] \\
HNC & $\nu_{1}$ & 226.9 &  3626.4--3629.1 & [a] \\
\enddata
\tablenotetext{}{[a] \citet{BOTSCHWINA1995345} }
\end{deluxetable}

\section{Sample kinetic traces for cyanobenzene photoproducts}\label{sec:time-products}

The temporal behavior of the minor photoproducts observed during the 3-hour 193 nm irradiation of a $\sim$40 ppm sample of cyanobenzene in p-H$_2$, as listed in Table \ref{tab:minor_products}. The kinetic traces are produced by integrating the IR bands with peaks at the wavenumbers indicated in the figure legend.

\begin{figure}[hbt!]
\centering
\includegraphics[width=0.6\textwidth]{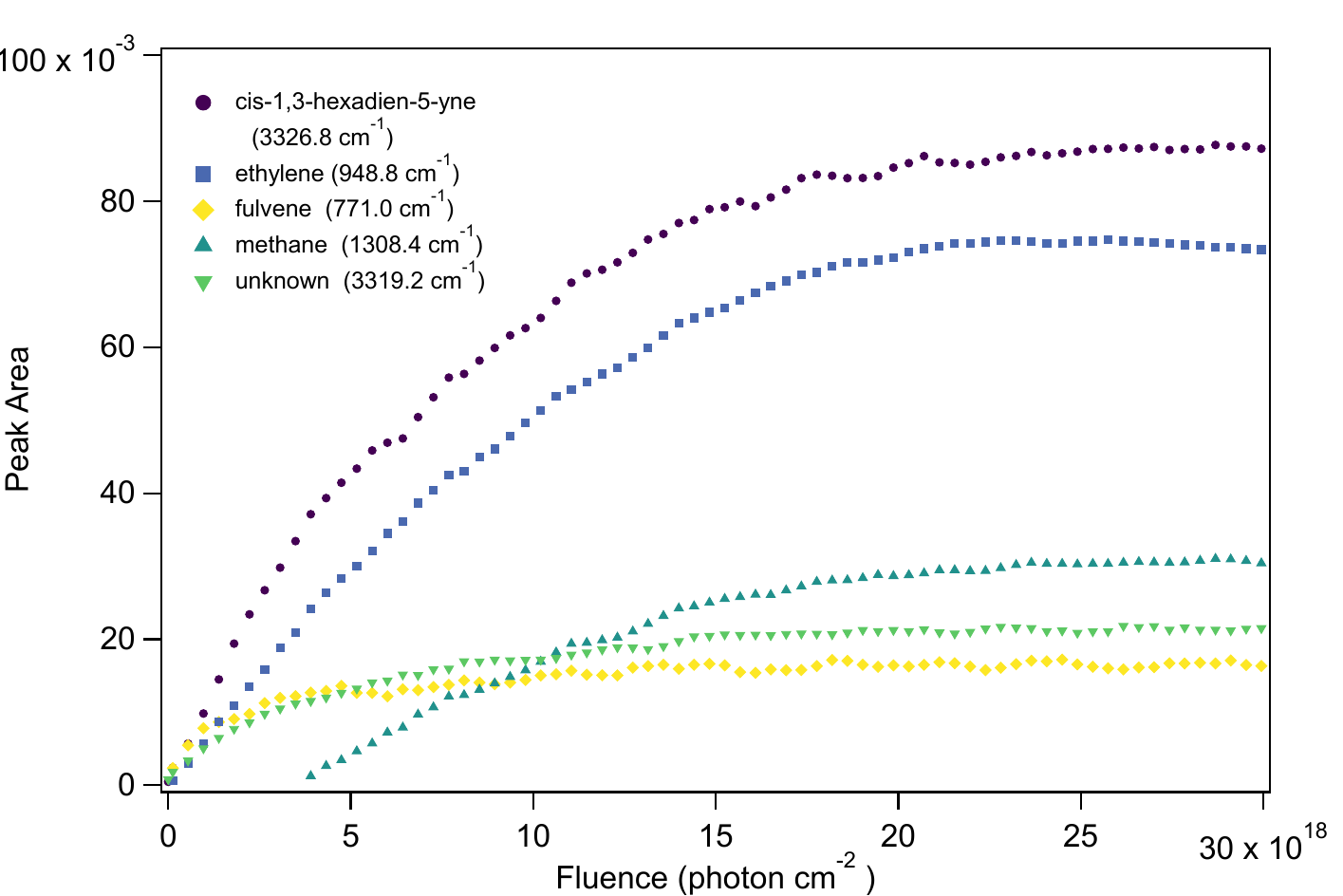}
\caption{Temporal behavior of the additional photoproducts listed in Table 3 that were observed to form during the irradiation of cyanobenzene at 193 nm}
\label{fig:photoproducts}
\end{figure}

\end{document}